\documentclass[12pt]{article}
\usepackage[utf8]{inputenc}
\usepackage[margin=1.3in]{geometry}

\usepackage[T1]{fontenc}
\usepackage{kpfonts,baskervald}

\usepackage{soul}
\usepackage{cite}

\usepackage{amsfonts}
\usepackage{mathrsfs}
\usepackage{amsmath,amssymb}
\usepackage{stmaryrd}
\usepackage{wasysym}

\usepackage[table,xcdraw]{xcolor}
\usepackage{enumitem}
\usepackage{booktabs}

\usepackage[all]{xy}

\usepackage{adjustbox}
\usepackage{tikz}
\usetikzlibrary{positioning}
\usetikzlibrary{cd}
\usetikzlibrary{decorations.markings}
\usepackage{pgfplots}
\usepackage[labelfont=bf]{caption}
\usepackage{blindtext}

\usepackage{amsthm}
\theoremstyle{definition}

\numberwithin{equation}{section}

\usepackage{hyperref}
\definecolor{airforceblue}{rgb}{0.36, 0.54, 0.66}
\definecolor{dgreen}{rgb}{0.09, 0.45, 0.27}
\hypersetup{
    colorlinks,
    citecolor=purple,
    filecolor=black,
    linkcolor=airforceblue,
    urlcolor=blue
}
\usepackage[nameinlink]{cleveref}

\newcommand*\diff{\mathop{}\!\mathrm{d}}

\newcommand*\I{\mathrm{i}}

\newcommand*\cL{\mathcal{L}}
\newcommand*\cM{\mathcal{M}}

\newcommand*\cT{\mathcal{T}}

\newcommand*\bbR{\mathbb{R}}

\newcommand*\bbZ{\mathbb{Z}}

\newcommand*\U{\mathrm{U}}
\newcommand*\SU{\mathrm{SU}}
\newcommand*\PSU{\mathrm{PSU}}

\newcommand{\be}{\begin{equation}}
\newcommand{\ee}{\end{equation}}

\usepackage{comment}

\begin{document}

\begin{titlepage}
\thispagestyle{empty}
\phantom{fez}

\vspace{-0.5cm}

\begin{center}

{\Huge {\bf Remarks on Geometric Engineering, Symmetry TFTs and Anomalies}}

\vspace{0.5cm}

{\Large Michele Del Zotto,$^{\#}$ Shani Nadir Meynet,$^\circ$ \\ and Robert Moscrop$^\bullet$}

\vspace{1cm}

{\footnotesize

$^{\#,\circ,\bullet}$ \textit{Department of Mathematics, Uppsala University\\
Box 480, SE-75106 Uppsala, Sweden} \\

\medskip

$^{\#,\circ}$ \textit{Centre for Geometry and Physics, Uppsala University\\
Box 480, SE-75106 Uppsala, Sweden} \\

\medskip

$^{\#}$  \textit{Department of Physics and Astronomy, Uppsala University\\
Box 520, SE-75106 Uppsala, Sweden}

\medskip

$^{\bullet}$  \textit{Department of Physics, University of Cincinnati, 400 Geology/Physics Bldg\\
PO Box 210011, Cincinnati OH 45221, US}

\medskip

$^{\bullet}$  \textit{Center  of  Mathematical  Sciences  and  Applications,  Harvard  University,\\  MA  02138,  USA}

\medskip

}

\footnotesize{{\tt \href{mailto:michele.delzotto@math.uu.se}{michele.delzotto@math.uu.se}}}, \footnotesize{{\tt \href{mailto:shani.meynet@math.uu.se}{shani.meynet@math.uu.se}}}\\
\footnotesize{{\tt \href{mailto:robert@cmsa.fas.harvard.edu}{robert@cmsa.fas.harvard.edu}}}
\vspace{1cm}

{\large \textbf{Abstract}}

\end{center}

\noindent Geometric engineering is a collection of tools developed to establish dictionaries between local singularities in string theory and (supersymmetric) quantum fields. Extended operators and defects, as well as their higher quantum numbers captured by topological symmetries, can be encoded within geometric engineering dictionaries. In this paper we revisit and clarify aspects of these techniques, with special emphasis on 't Hooft anomalies, interpreted from the SymTFT perspective as obstructions to the existence of Neumann boundary conditions. These obstructions to gauging higher symmetries are captured via higher link correlators for the SymTFT on spheres. In this work, we give the geometric engineering counterpart of this construction in terms of higher links of topological membranes. We provide a consistency check in the context of 5D SCFTs with anomalous 1-form symmetries, where we give two independent derivations of the anomaly in terms of higher links, one purely field theoretical and the other purely geometrical. Along the way, we also recover the construction of non-invertible duality defects in 4D $\mathcal N=4$ SYM from a geometric engineering perspective.

\vfill

\begin{flushright}
\noindent $\underline{\quad\quad\quad\quad\quad\quad\quad\quad\quad\quad\quad\;\;}$\\

February, 2024

\end{flushright}

\end{titlepage}


\setcounter{page}{1}
{\hypersetup{linkcolor=black}
\tableofcontents
}
\section{Introduction}

Following the seminal work of Gaiotto, Kapustin, Seiberg and Willet \cite{Gaiotto:2014kfa}, a growing amount of evidence indicates that symmetries of quantum fields are encoded in collections of topological defects and operators. In particular, generalised conserved quantum numbers of extended operators and defects are captured in this way -- see \cite{Cordova:2022ruw,McGreevy:2022oyu,Freed:2022iao,Gomes:2023ahz,Schafer-Nameki:2023jdn,Brennan:2023mmt,Bhardwaj:2023kri,Shao:2023gho,Carqueville:2023jhb} for some recent reviews. 

There are various approaches so far to capture the topological subsectors of a given QFT. Often one can start from a Lagrangian description of the theory and explicitly construct the generalised symmetries of interest \cite{Aharony:2013hda,
Kapustin:2013uxa,
Kapustin:2014gua,
Kapustin:2014zva,
Yoshida:2015cia,
Thorngren:2015gtw,
Tachikawa:2017gyf,
Cordova:2018cvg,
Benini:2018reh,
Rudelius:2020orz,
Hidaka:2020iaz,
Hsin:2020nts,
Cordova:2020tij,
BenettiGenolini:2020doj,
Hidaka:2020izy,
Brennan:2020ehu,
Heidenreich:2020pkc,
Hidaka:2021kkf,
Genolini:2022mpi,
Damia:2022seq,
Delmastro:2022pfo}. Another option is to formulate the theory on the lattice when possible and then realise the symmetry operators as topological defects defined by actions on links and sites \cite{Aasen:2016dop,Freed:2018cec,Aasen:2020jwb,Inamura:2021szw,Bochniak:2021ibv, Abe:2022nfq, 
Delcamp:2022sjf,Koide:2023rqd, Kan:2023yhz, Abe:2023uan, Abe:2023ubg,Inamura:2023qzl,Seiberg:2023cdc, Abe:2023ncy, Sinha:2023hum,Abe:2023atv,Honda:2024yte,Seiberg:2024gek
}. A third option is to exploit an auxiliary topological theory defined in one higher-dimension, the \textit{topological symmetry theory} or \textit{SymTFT} for short \cite{Ji:2019jhk,Gaiotto:2020iye,
Apruzzi:2021nmk,
Freed:2022qnc,
Kaidi:2022cpf,
Kaidi:2023maf,
Bhardwaj:2023wzd,
Baume:2023kkf,
Brennan:2024fgj,
Antinucci:2024zjp,
Bonetti:2024cjk
}. 
Among the most interesting features of generalised symmetries is that the corresponding topological defects are often non-invertible \cite{Bhardwaj:2017xup,Chang:2018iay,Komargodski:2020mxz,Nguyen:2021naa,Heidenreich:2021xpr,Thorngren:2021yso,Sharpe:2021srf,
Huang:2021zvu,
Choi:2021kmx,
Kaidi:2021xfk,
Roumpedakis:2022aik,
Bhardwaj:2022yxj,
Arias-Tamargo:2022nlf,
Choi:2022zal,
Kaidi:2022uux,
Choi:2022jqy,
Cordova:2022ieu,
Antinucci:2022eat,
Damia:2022bcd,
Chang:2022hud,
Choi:2022rfe,
Bhardwaj:2022lsg,
Bartsch:2022mpm,
Niro:2022ctq,
Chen:2022cyw,
Karasik:2022kkq,
Decoppet:2022dnz,
GarciaEtxebarria:2022jky,
Bhardwaj:2022kot,
Bhardwaj:2022maz,
Bartsch:2022ytj,
Hsin:2022heo,
Zhang:2023wlu,
Choi:2023xjw,
Anber:2023pny,
Bhardwaj:2023ayw,
Bartsch:2023wvv,
Copetti:2023mcq,
Decoppet:2023bay,
Chen:2023czk,
Sun:2023xxv,
Cordova:2023bja,
Antinucci:2023ezl,
Choi:2023vgk,
Diatlyk:2023fwf,
Nagoya:2023zky,
Perez-Lona:2023djo,
Sela:2024okz,DZGH
}, actually exhibiting a higher structure\footnote{\ See \cite{Copetti:2023mcq} for a work where the first layer of the higher structure of the chiral symmetry of massless QED (the F-symbols) are characterised and used to constrain correlators and partition functions.}  --- see \cite{Gaiotto:2017yup,
Hsin:2018vcg,Garcia-Etxebarria:2018ajm,
Buican:2021uyp,
Hayashi:2022oxp,
Kaya:2022edp,
Cordova:2022rer,
DelZotto:2022ras,
Hayashi:2022fkw,
Bhardwaj:2022dyt,
Lin:2022xod,
Cordova:2022fhg,
Choi:2022fgx,
Yokokura:2022alv,
Giaccari:2022xgs,
Cordova:2022qtz, 
Apte:2022xtu, 
Lin:2023uvm,
Putrov:2023jqi,
Damia:2023ses,
Argurio:2023lwl,
vanBeest:2023dbu,
vanBeest:2023mbs,
Cordova:2023ent,
Pace:2023kyi,
Choi:2023pdp,
Cordova:2023her,
Damia:2023gtc,
Bhardwaj:2023idu,
Bhardwaj:2023fca,
Brennan:2023kpo,
Brennan:2023vsa,
Cordova:2023qei,
Bhardwaj:2023bbf,
Brennan:2023ynm,
Dumitrescu:2023hbe,
Cordova:2023jip, 
Cordova:2024ypu} for an (incomplete) list of recent applications to QFTs in various dimensions.

It is believed, based on the landscape of supersymmetric examples, that the vast majority of quantum fields lack a conventional Lagrangian description. In order to characterise the corresponding generalised symmetries and the SymTFTs for these systems new tools are necessary. A powerful technique is to exploit the higher dimensional SCFT origin of lower dimensional systems \cite{Tachikawa:2013hya,Gukov:2020btk,Bashmakov:2022jtl,Bashmakov:2022uek,Antinucci:2022cdi,Bashmakov:2023kwo,Chen:2023qnv}. Another strategy consists in realising the symmetries of the field theories of interest from geometric engineering \cite{DelZotto:2015isa,
Albertini:2020mdx,
Morrison:2020ool,GarciaEtxebarria:2019caf,DelZotto:2020esg,
Apruzzi:2021vcu,
Hosseini:2021ged,
Closset:2021lhd,
Bhardwaj:2021wif,
Apruzzi:2021mlh,
Bhardwaj:2021mzl,
Hubner:2022kxr,
DelZotto:2022joo,
DelZotto:2022fnw,
Apruzzi:2022dlm,
Cvetic:2022imb,
Heckman:2022muc,
Mekareeya:2022spm,
vanBeest:2022fss,
Heckman:2022xgu,
Amariti:2023hev,
Carta:2023bqn,
Acharya:2023bth,
Dierigl:2023jdp,
Cvetic:2023plv,
Lawrie:2023tdz,
Apruzzi:2023uma,
Closset:2023pmc,
Yu:2023nyn,
Garding:2023unh,
Heckman:2024obe,Closset:2020scj,Closset:2021lwy
}  or via holography \cite{Aharony:1998qu,Witten:1998wy,Hofman:2017vwr,Apruzzi:2021phx,Bergman:2022otk,
GarciaEtxebarria:2022vzq,Apruzzi:2022rei,Antinucci:2022vyk,Apruzzi:2022nax,Etheredge:2023ler,Bah:2023ymy,Heckman:2024oot}. The main aim of this work is to clarify some features of this paradigm, especially highlighting the role of higher-link invariants and their interplay with anomalies.\footnote{\ We stress that the geometric engineering approach and the holographic one, albeit similar, still have some profound differences -- the results discussed in this note apply specifically to the geometric engineering case.} Consider a $(d+D)$-geometric engineering (GE) dictionary
\be\label{eq:geomenco}
X^d \quad \xrightarrow{\,\text{GE}\,} \quad \mathcal T_X \in \textsf{SQFT}_D
\ee
between $d$-dimensional singularities and $D$-dimensional supersymmetric field theories.\footnote{\ Here $d+D = 10, 11$ or $12$ according to whether we are considering superstring theories, M-theory or F-theory.} Recent progress indicates that $\mathcal T_X$ is always defined relative to a bulk $(D+1)$-dimensional theory, that in this paper we denote $\mathcal F_X$. In this construction $\mathcal F_X$ is supported on a geometry $\mathbb R_{\leq 0} \times M^D$, and $\mathcal T_X$ is located at the boundary $\{0\} \times M^D$. The theory $\mathcal F_X$ is obtained out of a geometric engineering limit `at infinity' on $\partial X$
\be\label{eq:geomencoinfo}
\partial X^{d-1} \quad \xrightarrow{\,\text{GE}^\infty\,} \quad \mathcal F_X \in \textsf{QFT}_{D+1}\,,
\ee
From this perspective the dictionaries in equation \eqref{eq:geomenco} needs to be supplemented by \eqref{eq:geomencoinfo}. In particular, depending on the geometry of $X$ and $\partial X$, the resulting bulk $D+1$ dimensional theory $\mathcal F_X$ can be a free theory, an interacting theory, a topological field theory, and very often it is a combination of various sectors of these types. Since geometric engineering backgrounds are non-compact, one must specify boundary conditions at infinity along $\partial X$. These are in turn interpreted as possible boundary conditions for the $(D+1)$-dimensional theory $\mathcal F_X$ at $\{-\infty\} \times M^D$. If the bulk theory $\mathcal F_X$ is topological and admits a topological boundary condition, $\mathcal B$, then this gives an isomorphism between the bulk-boundary system so obtained and a $D$-dimensional field theory $\mathcal T_X^\mathcal B$.\footnote{\ This idea was already hinted in the analysis of \cite{GarciaEtxebarria:2019caf,Albertini:2020mdx} building on the Freed-Moore-Segal (FMS) non-commuting fluxes \cite{Freed:2006yc,Freed:2006ya}. Interestingly, often it is not possible to fully specify such topological boundary conditions, this is the case for instance of IIB on certain ALE singularities, that give rise to 6d (2,0) SCFTs that are always defined relative to a 7D bulk.} Field theories obtained in this way typically have identical local operator spectrum and differ by their global structure.

In this paper, we focus mostly on applications of the engineering geometry to recover the topological sector of $\mathcal F_X$ which we denote $\mathcal F_X^{top}$. The topological operators of $\mathcal F_X^{top}$ are realised by membranes wrapping torsional cycles in $\partial X$. The main novelty in our approach is that we exploit the structure of linked membranes both in the $(D+1)$-dimensional bulk and in the $(d-1)$-dimensional boundary simultaneously to recover non-trivial specific higher-link correlators of $\mathcal F_X^{top}$ on $S^{D+1}$. Our main motivation to study these specific correlators is provided by their relation with 't Hooft anomalies, interpreted as obstructions to the existence of specific boundary conditions (Neumann as opposed to Dirichlet). As discussed recently by Kaidi, Nardoni, Zafir and Zheng \cite{Kaidi:2023maf}, the non-vanishing of one such higher-link correlator is the hallmark of such an anomaly. Our main aim in this work is to give the geometric engineering counterpart of this statement. We stress that while a Lagrangian description for the bulk $(D+1)$-dimensional symmetry theory is ambiguous, due to the many possible dualities leading to equivalent models, these correlators are not. An advantage of our proposal is that the whole geometric engineering geometry appears democratically, with higher-link invariants for the boundary of the engineering geometry and higher links in the bulk being captured by configurations of membranes, reminiscent of their higher $L/A_\infty$ structures and corresponding higher Massey products.\footnote{\ See eg. \cite{Sati:2008eg,Sati:2009ic,Fiorenza:2012ec} for some pioneering works in this direction in the context of string compactifications, we believe the structure we find here are the geometric engineering counterparts of these constructions.} In particular, we capture anomalies for generalised symmetries in terms of higher-link invariants.\footnote{\ In this work we ignore non-topological sectors of $\mathcal F_X$ and we focus on the subsector of $\mathcal F_X^{top}$ corresponding to finite generalized symmetries. Moreover, we also neglect more subtle aspects of the geometric engineering at infinity corresponding to the possible dependence on framing, which we believe corresponds to finer details of the SymTFT, such as Frobenius Schur indicators and more subtle fractionalization classes. We will refine our analysis in these directions in future works.}

Along the way we  discuss some features of finite 0-form symmetries in geometric engineering, which are realised on a slightly different footing than higher finite $p$-form symmetries. We propose two possible ways of obtaining such symmetries, one is based on the discrete isometries of $X$ (see e.g. \cite{Acharya:2021jsp} for a recent discussion), the other is based on stringy dualities. By combining these approaches, we give a geometric engineering description of the non-invertible duality defects of 4D $\mathcal N=4$ SYM from the perspective of Type IIA superstrings. Moreover, as further consistency checks for our proposal, we exploit the geometric engineering realisation of the 5D $\SU(p)_q$ SCFTs in M-theory, that have a well-known anomalous 1-form symmetry \cite{Gukov:2020btk}. In that context, we present a detailed field theory analysis that matches our geometric results with the corresponding 6d SymTFTs. In this example, we find that the anomalous topological defects are non-invertible in the bulk SymTFT.

Finally, our work further clarifies the role of the defect groups \cite{DelZotto:2015isa} (see also \cite{Albertini:2020mdx,Morrison:2020ool,GarciaEtxebarria:2019caf,Hubner:2022kxr}) in geometric engineering dictionaries. The defect groups are Pontryagin dual labels for the topological defects in the SymTFT, and indeed capture the possible configurations of membranes wrapped at infinity. The information captured by defect groups, however, needs to be supplemented by the data of 't Hooft anomalies, as possible obstructions to gauging outlined in \cite{Kaidi:2023maf}. The possible topological boundary conditions for $\mathcal F^{top}_X$ are severely constrained by this further input.\footnote{\ We thank Jonathan Heckman for a discussion about this idea at the Nordita Program \textit{Categorical Aspects of Symmetries} in August 2023, where some of the results discussed in this note have been presented.}

\bigskip

This paper is organised as follows. In Section \ref{sec:GE} we give a brief review of some features of geometric engineering dictionaries we will use in this work. In particular, we revist the action of finite 0-form symmetries arising from isometries and dualities and their interplay with defect groups. Section \ref{sec:linkings} contains the core of our proposal, in particular in \ref{sec:higherlinks} we propose a dictionary to capture correlators of the SymTFT on the sphere in terms of topological membrane higher links (Equation \eqref{eq:branelink}). In Section \ref{sec:examples} we discuss some examples to bring our methods to tests. Firstly in Section \ref{sec:7d} we recover the known features of 7D SYM theories as a warmup. We proceed in Section \ref{sec:N=4revisited} where we give the geometric engineering counterparts of the simplest non-invertible duality defects of 4D $\mathcal N=4$ SYM theories (after \cite{Kaidi:2021xfk,Choi:2021kmx,Kaidi:2022uux}) via type IIA superstrings. Finally, in Section \ref{sec:5d} we give a detailed discussion of the 5D SCFTs with gauge theory phases $\SU(p)_q$ where we recover from our formalism the results presented in \cite{Gukov:2020btk,BenettiGenolini:2020doj,Apruzzi:2021nmk}. All these examples have been selected because known results about these models can be obtained from purely field theoretical methods and we can use them as realiable testing grounds for our techniques. In Section \ref{sec:conclusions} we present some conclusions and some directions for future work building on this study. Appendix \ref{app:torsionallinks} contains a brief summary of the features of the higher links we consider as well as some further technicalities.

\section{Symmetries from geometry: a lightning review}\label{sec:GE}

In this section, to fix notations and conventions, we discuss the general framework of our study, giving a brief overview of the geometric engineering dictionaries \cite{Katz:1996fh, Katz:1996th, Bershadsky:1996nh}, emphasising the dependence on global structures and defect groups \cite{DelZotto:2015isa,Garcia-Etxebarria:2019caf, Albertini:2020mdx,Morrison:2020ool}. The main message is that the geometric engineering procedure establishes a dictionary between $d$-dimensional singular geometries and $D$-dimensional relative field theories, a nomenclature introduced in \cite{Freed:2012bs} to describe bulk-boundary systems. In this review we clarify some aspects of the geometric engineering dictionaries. In particular, in Section \ref{sec:hilb} we describe two alternative ways of obtaining examples of finite 0-form symmetries in geometric engineering: one arising purely from geometry (finite isometries), while the other, more quantum in nature, arises from stringy dualities.\footnote{\ The case of continuous global symmetries requires a different analysis -- see e.g. \cite{DelZotto:2014hpa,Hayashi:2019fsa,Eckhard:2020jyr,Acharya:2023bth,DeMarco:2023irn}. Some details about continuous non-abelian global symmetries can be found in \cite{Bonetti:2024cjk,DMDZM}.} In Section \ref{sec:defhilb}, we discuss the geometric origin of defect Hilbert spaces. Finally, in Section \ref{sec:rel}, we review the interplay between defect groups and global structures, clarifying their interpretation in terms of relative field theory \cite{Apruzzi:2021nmk,Apruzzi:2023uma}. 

\subsection{Hilbert space and 0-form symmetries from geometry}\label{sec:hilb}

Let $\mathscr S$ denote one out of the possible string theories, M-theory or F-theory. In this paper, we refer to geometric engineering as a collection of techniques establishing a correspondence between local stable $d$-dimensional BPS backgrounds $X$ for $\mathscr S$ and a $D$-dimensional quantum field theory $\mathcal T_X(\cdot)$. Schematically, we write this as
\begin{equation}
\mathcal T_X(\cdot) = GE_{\mathscr S / X}(\cdot).
\end{equation}
In geometric engineering, the (possibly twisted) partition function of $\mathcal T_X$ on a curved space $M$ is given by\footnote{\ In this note, we will often borrow the notation of functorial field theory, which suggests to interpret QFTs as higher functors from suitably enriched higher bordism categories to the category of vector spaces (see e.g. \cite{Tachikawa:2017byo,freedTFTbook} for reviews). In particular $\mathcal T(M)$ where $M$ is a $d$-dimensional closed compact manifold is our (schematic) notation for the partition function of $\mathcal T$ on $M$.}
\begin{equation}
\mathcal T_X(M) = GE_{\mathscr S / X}(M) \equiv \mathcal Z_{\mathscr S}(M\rtimes X),
\end{equation}
where on the RHS we are considering the partition function of $\mathscr S$ on the non-compact (possibly twisted) background $M\rtimes X$.

In this paper we wish to obtain $D$-dimensional SCFTs with a sufficient amount of supersymmetry such that the quantum stringy corrections to classical geometry are under control. Therefore the stable BPS backgrounds of interest in this paper are typically non-compact spaces $X$ with a complete metric $g$ of special holonomy that admit a singular limit at finite distance in moduli space where all scales in the geometry are sent to zero. We assume this is the case from now on.\footnote{\ We stress however that geometric engineering techniques can be extended outside of this realm.} Since $X$ is non-compact, the resulting theory depends on choices of boundary conditions at infinity for $\mathscr S$ along $\partial X$. We will return to this point below.

Proceeding to the next layer, we want to consider the Hilbert space that theory $\mathcal T_X$ assigns to a $(D-1)$-dimensional manifold $N$ in canonical quantization, denoted $\mathcal T_X(N)$. Here on we assume that $N$ is a compact spin manifold without torsion, to simplify our analysis. The Hilbert space $\mathcal T_X(N)$ is computed via supersymmetric quantum mechanics (SQM) quantizing configurations of membranes wrapping the compact cycles of $N \rtimes X$ and extended along $\mathbb R_\text{time}$
\be\label{eq:hilbert}
GE_{\mathscr S / N \rtimes X}(\mathbb R_\text{time})
\ee

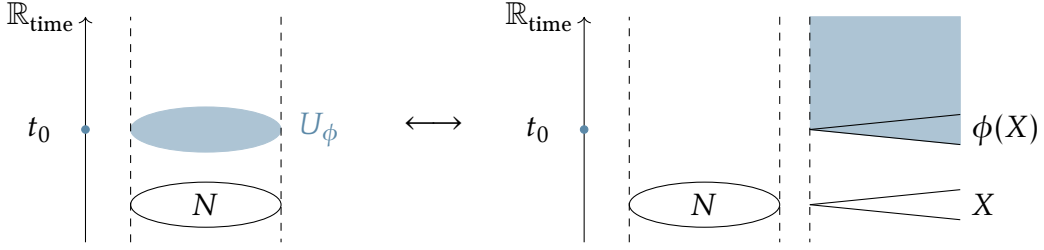
\begin{figure} 
\begin{center}
\begin{tabular}{p{5cm}cp{7cm}} 
$\begin{gathered}
\begin{tikzpicture}
    \draw [->] (0,0) -- (0,3);
        \node[left] at (0,3){$ \mathbb R_\text{time}$};
        \draw[fill,airforceblue] (0,1.5) circle (0.05cm);
        \node[left] at (-0.3,1.5){$t_0$};
        \node[airforceblue] at (3.1,1.5){$U_\phi$};
        \draw[fill,color=airforceblue!50] (1.6,1.5) ellipse (1cm and 0.3cm);
    \draw (1.6,0.5) ellipse (1cm and 0.3cm);
    \node at (1.6,0.5){$N$};
    \draw[dashed] (0.6,0) -- (0.6,3);
    \draw[dashed] (2.6,3) -- (2.6,0);
\end{tikzpicture}
\end{gathered}
$
&$\longleftrightarrow$&
$\begin{gathered}\begin{tikzpicture}
    \draw [->] (0,0) -- (0,3);
        \node[left] at (0,3){$ \mathbb R_\text{time}$};
                \draw[fill,airforceblue] (0,1.5) circle (0.05cm);
        \node[left] at (-0.3,1.5){$t_0$};
         \draw (1.6,0.5) ellipse (1cm and 0.3cm);
    \node at (1.6,0.5){$N$};
    \draw[dashed] (0.6,0) -- (0.6,3);
    \draw[dashed] (2.6,3) -- (2.6,0);
    \draw (3,0.5) -- (5,0.7);
    \draw (3,0.5) -- (5,0.3);
    \node[right] at (5,0.5){$X$};
    \draw[fill,color=airforceblue!50] (3,1.5) -- (5,1.3) -- (5,3) -- (3,3);
    \node[right] at (5,1.5){$\phi(X)$};
    \draw (3,1.5) -- (5,1.7);
    \draw (3,1.5) -- (5,1.3);
    \draw[dashed] (3,3) -- (3,0);
\end{tikzpicture}
\end{gathered}
$\\
\end{tabular}
\end{center}

\caption{(left) Action of a 0-form symmetry $U_\phi$ on the Hilbert space corresponding to $N$ vs. (right) The geometry $U(\phi,t_0)$ of equation \eqref{eq:zeroformaction}.}\label{fig:zeroform}
\end{figure}

It is interesting at this point to consider the geometric counterpart of the action of 0-form symmetries on the Hilbert space. In what follows we will focus on the case of finite 0-form symmetries. Recall that a 0-form symmetry for a $D$-dimensional theory corresponds to a topological operator of codimension 1, let us call such an operator $U_\phi$. Consider acting on the Hilbert space $\mathcal T_X(N)$ with an insertion of $U_\phi$ at $t = t_0$ along $N$, as shown in Figure \ref{fig:zeroform} (left). It is easy to realise such an action via geometric engineering exploiting the group $\mathscr I(X,g)$ of isometries of $(X,g)$ -- see e.g. \cite{Acharya:2021jsp}. This realises very explicitly a subgroup of the 0-form symmetries of $\mathcal T_X$. To engineer the action of such 0-form symmetries on the Hilbert space geometrically, we can proceed as follows. Pick an element $\phi \in \mathscr I(X,g)$ and consider the fibration over $\mathbb R_{\text{time}}$ on the RHS of Figure \ref{fig:zeroform}
\be\label{eq:zeroformaction}
U(\phi,t_0) = \begin{cases} N \rtimes X \times \{t\}, &\text{for } t \in \mathbb R_\text{time},\, \, t < t_0,\\  N \rtimes \phi(X)\times \{t\},\ &\text{for } t \in \mathbb R_\text{time},\, \, t \geq t_0.\end{cases}
\ee 
Since wrapped membranes of $\mathscr S$ give rise to the states in the Hilbert space $\mathcal T_X(N)$ and, in the cases of interest in this paper, singular homology with integer coefficients captures faithfully the relevant membrane charges, the action of the corresponding operator $U_\phi$ on the states in the Hilbert space is induced from the push forward action in homology 
\be
\phi_*: H_\bullet(X,\mathbb Z) \to H_\bullet(X,\mathbb Z).
\ee
We stress here that often from the geometric engineering perspective the various moduli of the geometry correspond to both parameters and vevs, and therefore geometric engineering puts duality symmetries and accidental symmetries on the same footing. For this reason isometries can give rise to an accidental symmetry (if the enhancement occurs at points in moduli space corresponding to tuning of vevs in the geometric engineering dictionary) or to a duality symmetry (if the enhancement occurs by tuning moduli corresponding to parameters in the geometric engineering dictionary). It is also possible that mixed configurations occur, namely that the enhanced symmetry arises at points in moduli space corresponding to tuning both parameters and vevs. In this case the symmetry is an accidental duality symmetry.\footnote{\ The vast majority of 5D dualities from this perspective can give rise to accidental duality symmetries, as these require tuning both relevant deformations of the SCFT and vevs at special values\cite{Aharony:1997bh,Closset:2018bjz,Bhardwaj:2019ngx,Closset:2020afy} -- more details about this are discussed in \cite{DMDZM}.}

On top of 0-form symmetries from isometries, there is another class of symmetries which arise from quantum effects in string theory. As an example, consider a stringy duality that relates $\mathscr S$ on the geometry $X$ to $\mathscr S$ on a geometry $S(X)$. In general this does not correspond to a non-trivial automorphism of the theory $\mathcal T_X$. However, it can happen that under special circumstances, such as the existence of a self-dual point in moduli space, the following holds
\be
S(X_*) = X_* .
\ee
In this case the stringy duality corresponds to a non-trivial automorphism of the theory $\mathcal T_{X_*}$. The symmetry is engineered as in Figure \ref{fig:zeroform}, replacing $\phi$ with the stringy duality operation $S$ such that $S(X_*) = X_*$. Again, $S$ acts on the spectrum of wrapped branes (and thus on the Hilbert space) as dictated by the action of the duality. As an example, consider IIA and IIB superstrings and their self T-duality upon volume inversion of a two-dimensional torus $T^2$: if a given geometry has a torus fibration such that the volume of the torus is a parameter, then that geometry can give a symmetry enhancement at the self-dual volume.  We will discuss an an explicit example of this effect in details below in Section \ref{sec:N=4revisited}.

\medskip

There is of course a \textit{caveat} in the above discussion. The 0-form symmetries we discussed above are often just a subgroup of the total 0-form symmetries of $\mathcal T_X$. In most of our examples, $X$ is a Calabi-Yau threefold singularity and in those cases the above construction typically realises examples of finite 0-form symmetries of $\mathcal T_X$. The case of continuous 0-form symmetries is very different \cite{Hayashi:2019fsa,Eckhard:2020jyr,DelZotto:2022joo,Acharya:2023bth,Heckman:2024oot,Bonetti:2024cjk}.

\subsection{Defect Hilbert spaces and defect groups}\label{sec:defhilb}
In order to construct the Hilbert space of $\mathcal T_X$ in the presence of a probe defect operator of a given charge, one has to consider quantizing the theory with an insertion of such a defect along the time direction. In geometric engineering bare defects of dimension $(p-k+1)$ are obtained by choosing a non-compact $k$-dimensional cycle 
\begin{gather}
    S \in H_k(X,\partial X)
\end{gather}
and considering a $p$-brane $\textbf{Mp}$ wrapped on it. Since the cycle is non-compact, the resulting configuration has infinite energy, as expected of a defect. We denote such a probe defect by
\begin{gather}\label{eq:wilson} 
     \mathcal W_\textbf{Mp}^S.
\end{gather}
Similarly, we denote by
\be
\mathcal T_X(\mathcal W_\textbf{Mp}^S(\gamma) \subset M)\,
\ee 
the partition function on a spacetime $M$ with defect $\mathcal W_\textbf{Mp}^S$ inserted along a $(p-k+1)$-dimensional  cycle $\gamma \subset M$. These are encoded by geometric engineering through the relation
\be
\mathcal T_X(\mathcal W_\textbf{Mp}^S(\gamma) \subset M) = \mathcal Z_{\mathscr S}(\textbf{Mp}(\gamma \rtimes S) \subset M\rtimes X)\,,
\ee
where on the RHS we are considering the partition function of $\mathscr S$ theory on a background $M\rtimes X$ in the presence of the insertion of an $\textbf{Mp}$ brane on $\gamma \rtimes S \subset M \rtimes X$.

Another interesting quantity that one can compute is the so called \textit{defect Hilbert space}, which is obtained by considering a $D$-dimensional spacetime of the form $\mathbb R_{\text{time}} \times N$, as above, and inserting $\mathcal W_\textbf{Mp}^S$ along the time direction times a $(p-k)$-dimensional compact submanifold $\Sigma\subset N$. We denote by 
\be
\mathcal T_X(\mathcal W_\textbf{Mp}^S(\Sigma) \subset N)
\ee
the corresponding defect Hilbert space. In geometric engineering this is realised as follows. The membrane $\textbf{Mp}$ is inserted along $\mathbb R_\text{time} \times \Sigma \rtimes S$ in the background $\mathbb R_\text{time} \times N \rtimes X$, and the procedure described around equation \eqref{eq:hilbert} is applied in presence of such an additional wrapped membrane, thus changing the corresponding supersymmetric quantum mechanical models that encode the defect Hilbert space:
 \be
\mathcal T_X(\mathcal W_\textbf{Mp}^S(\Sigma) \subset N) = GE_{\mathscr S/\mathbf{Mp}(\Sigma \rtimes S) \subset N \rtimes X}(\mathbb R_{\text{time}}).
\ee
Notice that a defect Hilbert space so defined differs from the original Hilbert space of the theory in equation \eqref{eq:hilbert} since, on top of the states in the bulk Hilbert space, we have additional sectors arising from bound states between bulk excitations and the defect. These are captured by the bound states between the wrapped $\textbf{Mp}$ brane on $\mathbb R_{\text{time}} \times \Sigma \rtimes S$ and the other membranes wrapped on (vanishing) cycles with finite volumes.
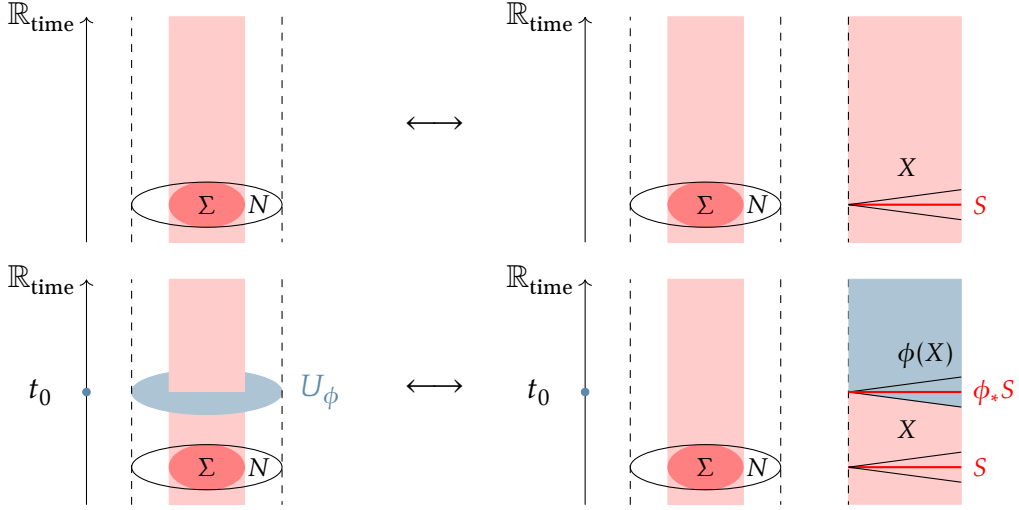
\begin{figure} 
\begin{center}
\begin{tabular}{p{5cm}cp{7cm}} 
$\begin{gathered}
\begin{tikzpicture}
    \draw [->] (0,0) -- (0,3);
        \node[left] at (0,3){$ \mathbb R_\text{time}$};
        \draw[fill,color=red!20] (1.1,0) -- (1.1,3) -- (2.1,3)-- (2.1,0);
        \draw[fill,color=red!50] (1.6,0.5) ellipse (0.5 cm and 0.3cm);
                \node at (1.6,0.5){{\footnotesize$\Sigma$}};
    \draw (1.6,0.5) ellipse (1cm and 0.3cm);
    \node at (2.3,0.5){{\footnotesize $N$}};
    \draw[dashed] (0.6,0) -- (0.6,3);
    \draw[dashed] (2.6,3) -- (2.6,0);
\end{tikzpicture}
\end{gathered}
$
&$\longleftrightarrow$&
$\begin{gathered}\begin{tikzpicture}
      \draw [->] (0,0) -- (0,3);
        \node[left] at (0,3){$ \mathbb R_\text{time}$};
        \draw[fill,color=red!20] (1.1,0) -- (1.1,3) -- (2.1,3)-- (2.1,0);
        \draw[fill,color=red!50] (1.6,0.5) ellipse (0.5 cm and 0.3cm);
    \node at (1.6,0.5){{\footnotesize$\Sigma$}};       
    \draw[fill,color=red!20] (3.5,0) -- (3.5,3) -- (5,3)-- (5,0);
    \draw [thick,red] (3.5,0.5) -- (5,0.5);
    \draw[dashed] (0.6,0) -- (0.6,3);
    \draw[dashed] (2.6,3) -- (2.6,0);
    \draw (3.5,0.5) -- (5,0.7);
    \node at (2.3,0.5){{\footnotesize $N$}};
    \draw (1.6,0.5) ellipse (1cm and 0.3cm);
    \draw (3.5,0.5) -- (5,0.3);
        \node[right] at (4,1){{\footnotesize $X$}};
    \node[right] at (5,0.5){{\footnotesize $\textcolor{red}{S}$}};
    \draw[dashed] (3.5,3) -- (3.5,0);
\end{tikzpicture}
\end{gathered}
$\\
$\begin{gathered}
\begin{tikzpicture}
    \draw [->] (0,0) -- (0,3);
        \node[left] at (0,3){$ \mathbb R_\text{time}$};
         \draw[fill,airforceblue] (0,1.5) circle (0.05cm);
        \node[left] at (-0.3,1.5){$t_0$};
        \node[airforceblue] at (3.1,1.5){$U_\phi$};
        \draw[fill,color=red!20] (1.1,0) -- (1.1,1.3) -- (2.1,1.3)-- (2.1,0);
        \draw[fill,color=airforceblue!50] (1.6,1.5) ellipse (1cm and 0.3cm);
        \draw[fill,color=red!20] (1.1,1.5) -- (1.1,3) -- (2.1,3)-- (2.1,1.5);
        \draw[fill,color=red!50] (1.6,0.5) ellipse (0.5 cm and 0.3cm);
        \node at (1.6,0.5){{\footnotesize$\Sigma$}};
    \draw (1.6,0.5) ellipse (1cm and 0.3cm);
    \node at (2.3,0.5){{\footnotesize $N$}};
    \draw[dashed] (0.6,0) -- (0.6,3);
    \draw[dashed] (2.6,3) -- (2.6,0);
\end{tikzpicture}
\end{gathered}
$
&$\longleftrightarrow$&
$\begin{gathered}\begin{tikzpicture}
      \draw [->] (0,0) -- (0,3);
        \node[left] at (0,3){$ \mathbb R_\text{time}$};
         \draw[fill,airforceblue] (0,1.5) circle (0.05cm);
        \node[left] at (-0.3,1.5){$t_0$};
        \draw[fill,color=red!20] (1.1,0) -- (1.1,3) -- (2.1,3)-- (2.1,0);
        \draw[fill,color=red!50] (1.6,0.5) ellipse (0.5 cm and 0.3cm);
                \node at (1.6,0.5){{\footnotesize$\Sigma$}};
    \draw (1.6,0.5) ellipse (1cm and 0.3cm);
    \node at (2.3,0.5){{\footnotesize $N$}};
    \draw[dashed] (0.6,0) -- (0.6,3);
    \draw[dashed] (2.6,3) -- (2.6,0);  
            \draw[fill,color=red!20] (3.5,0) -- (3.5,1.5) -- (5,1.3)-- (5,0);
    \draw [thick,red] (3.5,0.5) -- (5,0.5);
    \draw (3.5,0.5) -- (5,0.3);
        \node[right] at (4,1){{\footnotesize $X$}};
    \node[right] at (5,0.5){{\footnotesize $\textcolor{red}{S}$}};
    \draw[dashed] (3.5,3) -- (3.5,0);
     \draw[fill,color=airforceblue!50] (3.5,1.5) -- (5,1.3) -- (5,3) -- (3.5,3);
    \draw (3.5,1.5) -- (5,1.7);
    \draw (3.5,1.5) -- (5,1.3);
    \draw (3.5,0.5) -- (5,0.7);
            \node[right] at (4,2){{\footnotesize $\phi(X)$}};
     \node[right] at (5,1.5){{\footnotesize $\textcolor{red}{\phi_*S}$}};
         \draw [thick,red] (3.5,1.5) -- (5,1.5);
\end{tikzpicture}
\end{gathered}
$\\
\end{tabular}
\end{center}

\caption{{\small TOP:} (left) Defect Hilbert space corresponding to inserting  $\mathcal W_\textbf{Mp}^S(\Sigma)$ vs. (right) its geometric engineering. {\small BOTTOM:} Insertion of a 0-form symmetry and resulting morphism between defect Hilbert spaces.}\label{fig:defhilbert}
\end{figure}
Since all possible $p$-branes wrapping compact cycles of $N\rtimes X$ contribute to the defect Hilbert space, physically, the latter will depend on the defect charge $S$ only via its equivalence class defined up to screening
\begin{equation}
[S] \in H_k(X,\partial X) / H_k(X).
\end{equation}
This is a generalised version of the 't Hooft screening mechanism: $[S] \neq 0$ is a necessary condition for the stability of the given defect. Defects with $[S] = 0$ are \textit{endable}, meaning that (at sufficiently high energies) there are operators in the spectrum that can be used to split the defect apart, causing its screening. This motivates the definition of the \textit{defect group} for $\mathcal T_X$ as  \cite{DelZotto:2015isa,Garcia-Etxebarria:2019caf,Albertini:2020mdx,Morrison:2020ool,Hubner:2022kxr}
\begin{equation}
\begin{aligned}\label{eq:defectgroup}
        \mathbb{D}_X := \bigoplus_n \mathbb{D}^{(n)}_X \quad \text{where}\ \  \mathbb{D}^{(n)}_X =  \bigoplus_{p\text{-branes}}\left(\bigoplus_{k \text{ s.t. } \newline p-k+1=n} \left(\frac{H_k(X , \partial X)}{H_{k}(X)}\right)\right).
\end{aligned}\end{equation}
Elements of $\mathbb{D}^{(n)}_X$ correspond to charges of $n$-dimensional defects that cannot be screened.

\medskip

Let us briefly comment on the finite 0-form symmetries in the classes we discussed above and their interplay with defects. Of course, the action of the symmetry is encoded in geometric engineering also in presence of defects. In the case of symmetries originating from isometries, the 0-form symmetry action on defects is obtained by lifting the isometry on the relative homology lattice. For the symmetries that have an origin via a self-duality transformation, their action is induced by the action of the duality on membrane charges. We will see an explicit example in Section \ref{sec:N=4revisited} below. Of course, the 0-form symmetry can act on higher dimensional defects as well by crossing (see e.g. the bottom part of Figure \ref{fig:defhilbert}). In the context of the defect Hilbert spaces, this has an interesting effect: if a given defect is charged with respect to the 0-form symmetry, one obtains a morphism $U_\phi(N)$ from one defect Hilbert space to the other:
\be
U_\phi(N) \in \text{Hom}\left(\mathcal T_X(\mathcal W_{\mathbf{Mp}}^{S}(\Sigma) \subset N),\mathcal T_X(\mathcal W_{\mathbf{Mp}}^{\phi_*S}(\Sigma) \subset N)\right)
.
\ee
However, it can happen that the two defects are equivalent in the defect group, meaning that 
\be
[\phi_*S] = [S] \in \mathbb D^{(n)}_X.
\ee
If this is the case, the morphism above can be interpreted as an automorphism, i.e. an invertible symmetry of the corresponding defect Hilbert space (the latter is expected to depend only on the class of $S$ in the defect group). Otherwise, if $[\phi_*S] \neq [S] \in \mathbb D^{(n)}_X$, then $\phi$ gives rise to a topological interface (a morphism) between two inequivalent defect Hilbert spaces. In this second case, this signals that a non-trivial representation of the 0-form symmetry is acting on the generators of the $n$-form symmetry.\footnote{\ This fact can be exploited to detect mixed 't Hooft anomalies generalising to this setup the mechanism discussed in \cite{Tachikawa:2013hya,Bashmakov:2022jtl} -- see \cite{DMDZM} for explicit examples.}

\medskip
 
We give a schematic description of these effects in Figure \ref{fig:defhilbert}. On the top of the Figure we represent the geometric engineering of a defect Hilbert space, while on the bottom of the figure we describe the geometric engineering of a morphism between the defect Hilbert spaces.

\subsection{Defect groups and relative theories}\label{sec:rel}

Because of the mutual non-locality of membranes in the theory $\mathscr S$ (as a consequence of electromagnetic duality and flux non-commutativity \cite{Freed:2006ya,Freed:2006yc}, and/or as a manifestation of more subtle effects related to the topology of the boundary), it can happen that the defects captured by the defect group are not all mutually local.  In these cases, the theory $\mathcal T_X$ it is understood as a \textit{relative field theory} \cite{Freed:2012bs,Tachikawa:2013hya}, namely $\mathcal T_X$ can be interpreted as a $D$-dimensional boundary coupled to a $(D+1)$-dimensional bulk theory, that we denote $\mathcal F_X$. The bulk $\mathcal F_X$ theory is placed on background $\mathbb R_{\leq 0} \times M$ where $M$ is the $D$-dimensional spacetime of interest. The bulk theory couples to the theory $\mathcal T_X$, which is placed on the boundary $\{0\} \times M$. Since $\mathcal F_X$ assigns to a $D$-manifold $M$ a Hilbert space $\mathcal F_X(M)$ and $\mathcal T_X$ is a boundary for $\mathcal F_X$, the value of $\mathcal T_X$ on a compact $D$-manifold $M$ is a vector in the Hilbert space $\mathcal T_X(M) \in \mathcal F_X(M)$, the so-called \textit{partition vector of  $\mathcal T_X$ on $M$}. In this way, the mutual non-locality of defects of $\mathcal T_X$ is no longer an issue-- collections of mutually local defects belong to distinct selection sectors in $\mathcal F_X(M)$.
 
From the perspective of geometric engineering, generalising remarks in \cite{GarciaEtxebarria:2019caf,Albertini:2020mdx}, we are lead to identify the Hilbert space $\mathcal F_X(M)$ with the geometric engineering Hilbert space that $\mathscr S$ assigns to the boundary of the spacetime, $\mathcal H_{\mathscr S}(\partial (M\times X))$. Assuming that $M$ is compact, spin and torsionless, and moreover that $\partial X$ has no singularities, one finds that $\mathcal H_{\mathscr S}(\partial (M\times X))$ is graded by the Pontryagin duals of the defect groups in equation \eqref{eq:defectgroup} by construction.\footnote{\ It is interesting, but beyond the scope of the present note, to extend this analysis further to the case where $M$ has torsion or is non-compact.} Indeed, these are realised in first approximation by exponentiated string theory fluxes measuring the membrane charges that contribute to the defect group, or equivalently via membranes wrapped at infinity (see e.g. \cite{Heckman:2022muc}). 

Focusing on the torsional parts of $\mathbb D_X$ corresponds to considering a subsector of the bulk theory which is typically a finite higher group gauge theory of generalised Dijkgraaf-Witten type (possibly with twists of various kinds).\footnote{\ This is a straightforward generalisation of the original definition of Dijkgraaf-Witten theory \cite{Dijkgraaf:1989pz} --- for a review see \cite{Sun:2023xxv}.} We call this topological subsector $\mathcal F^{top}_X(\cdot)$. 

\begin{figure}
    \centering
    \begin{tikzpicture}
        \shadedraw [shading=axis,top color = airforceblue, bottom color=white, white] (0,0) rectangle (5,3);
        \draw [airforceblue] (0,3) -- (5,3);
        \draw [very thick] (0,0) -- (0,3);
        \draw [very thick] (5,0) -- (5,3);
        \node[below] at (0,-0.1){$ \mathcal B$};
        \node[below] at (5,-0.1){$\mathcal T_X$};
        \node[above] at (2.5,1.2){$\mathcal F^{top}_X$};
        \draw [very thick] (10,0) -- (10,3);
        \draw [-stealth, thick] (6,1.5) -- (9,1.5);
        \node[above] at (7.5,1.5){contract};
        \node[below] at (10,-0.1){$\mathcal{T}^{\mathcal B}_X$};
    \end{tikzpicture}
    \caption{The `sandwich' picture. The leftmost boundary encodes a topological boundary $\mathcal B$ for the bulk $(D+1)$-dimensional theory $\mathcal F_X$. Being topological, we can stack the boundary $\mathcal B$ on the boundary supporting the theory $\mathcal T_X$. Contracting the interval, we obtain a $D$-dimensional field theory that has a partition function on compact torsionless $D$-manifolds.}\label{fig:sandwich}
\end{figure}
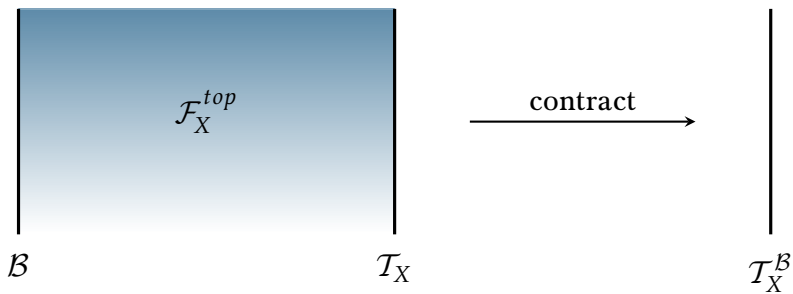

If the theory $\mathcal F^{top}_X(\cdot)$ admits a $D$-dimensional topological boundary $\mathcal B$ (i.e. a topological interface with the $(D+1)$-dimensional unit theory $\mathbf{1}_{D+1}$),\footnote{\ Or more generally an invertible SPT in presence of backgrounds.} this construction gives an example of the so-called \textit{topological symmetry theory} (or SymTFT)\cite{Kapustin:2014gua,Gaiotto:2020iye,Apruzzi:2021nmk,Freed:2022qnc}. When this is the case, giving boundary conditions at minus infinity on the semi-infite geometry $\mathbb R_{\leq 0} \times M$ is equivalent to give them at $-\epsilon$ on the compact geometry
\be
[-\epsilon,0] \times M \,,
\ee
for $\epsilon > 0$. In summary, we are inserting $\mathcal B$ and $\mathcal T_X$ on the opposite sides of the interval $[-\epsilon,0]$ -- see Figure \ref{fig:sandwich}. Since $\mathcal B$ is topological and so is $\mathcal F^{top}_X$, sending $\epsilon \to 0$ does not affect correlators on this geometry. In this way one obtains a theory $\mathcal T_X^{\mathcal B}$ that is no longer relative to the bulk $\mathcal F^{top}_X$ but depends on the choice of $\mathcal B$. In particular, all the topological membranes acting as generalised finite symmetries on $\mathcal T_X^{\mathcal B}$ can be read off from this construction, by extending bulk topological defects along $\mathcal B$.

From this perspective, theories obtained in this way generically won't have intrinsically non-invertible symmetries as a na\"ive application of the results of \cite{Kaidi:2022cpf}. We will shortly see that this expectation is too na\"ive: including stringy dualities in this formalism can be exploited to give rise to more subtle effects.\footnote{\ Other subtleties in the above dictionary arise when relaxing the condition that $\partial X$ is not singular. This is related to the case of continuous 0-form symmetries as discussed in \cite{Bonetti:2024cjk}.}

\medskip

Understanding the detailed properties of $\mathcal F^{top}_X$ from the perspective of geometric engineering  entails refining the previous description in \cite{Garcia-Etxebarria:2019caf,Albertini:2020mdx} that was based solely on flux non-commutativity. Indeed, from that perspective, one can identify the possible topological boundary conditions $\mathcal B$ via the maximally commuting subalgebras of the Heisenberg algebras of fluxes, but this approach can fail to capture obstructions to gauging that can arise from non-trivial higher links. The latter play an important role in the theory of generalised symmetries, as reviewed in detail in the recent paper by Kaidi, Nardoni, Zafrir and Zheng \cite{Kaidi:2023maf}. Several proposals have been advanced in the recent literature, mostly focusing on holographic systems (with a few exceptions), where a hybrid formalism is developed to describe the topological subsector of the $(D+1)$-dimensional bulk theory -- see e.g. \cite{Apruzzi:2021nmk,vanBeest:2022fss,Apruzzi:2022rei,GarciaEtxebarria:2022vzq,Heckman:2022xgu,Etheredge:2023ler,Apruzzi:2023uma,Bah:2023ymy,Baume:2023kkf}. In this work we revisit some aspects of this formalism, focusing solely on the geometric engineering case. Building on the recent works \cite{Heckman:2022muc,Baume:2023kkf}, we clarify a conjecture to uniformly capture the structure of the correlators of the topological membranes of $\mathcal F^{top}_X$ that encode these obstruction to gaugings in terms of vevs of specific higher links on $S^{D+1}$. In the next section we discuss some basic general features and in the rest of this paper we discuss some detailed examples.

\section{Topological defects and branes at infinity}\label{sec:linkings}

As we have reviewed in the previous section, geometric engineering assigns to a given singular geometry $X$ a theory $\mathcal T_X$, that in general is a relative field theory realised along a $D$-dimensional boundary of a $(D+1)$-dimensional bulk theory, $\mathcal F_X$, with a topological subsector $\mathcal F_X^{top}$.  In this section we discuss some general features of the topological defects and operators of $\mathcal F_X^{top}$. In  Section \ref{sec:generalremks} we review how these are realized from branes wrapped at infinity. In Section \ref{sec:higherlinks} we discuss how the higher obstructions to gauging detected by higher links are realised via geometric engineering at infinity.

\subsection{Relative SCFTs and Heisenberg algebras from geometry}\label{sec:generalremks}

To simplify the discussion of our examples, we assume that the theory $\mathcal T_X$ is a relative superconformal field theory (SCFT), obtained from geometric engineering on a $d$-dimensional conical singularity with special holonomy
\be
X = \mathrm{Cone}(\mathbf L_X),\, \qquad \diff s^2_X = \diff r^2 + r^2 \diff s^2_{\mathbf{L}_X}.
\ee
The space $\mathbf L_X$ is the \textit{link} of the singularity.\footnote{\ It is a $(d-1)$-dimensional compact manifold obtained by intersecting the given singularity with a sphere centered on it. If $X$ has special holonomy, the corresponding link has a metric with special properties inherited from the special holonomy of $X$. For example, if $X$ is a Calabi-Yau singularity with special holonomy $\SU(n)$, the resulting link is a $(2n-1)$-dimensional manifold with a Sasaki-Einstein metric (possibly with orbifold singularities).} For this class of examples, the structure of a bulk-boundary system is manifest in the geometric engineering formalism: the $(D+1)$-dimensional bulk theory, which we denote $\mathcal F_X$, arises from a geometric engineering `at infinity'
\begin{equation}
\mathcal F_X(\cdot) = GE^\infty_{\mathscr S/\mathbf{L}_X}(\cdot).
\end{equation}
Here the superscript $^\infty$ is meant to stress that since the space $\mathbf{L}_X$ is at infinite distance in the conical metric $\diff s^2_X = \diff r^2 + r^2 \diff s^2_{\mathbf L_X}$, its volume is effectively infinite as $r\to \infty$. For this reason, compactifying the string theory $\mathscr S$ on $\mathbf L_X$ at infinity requires care. In particular, we expect that all the dynamical degrees of freedom decouple and one is left with a theory $\mathcal F_X$ that has no relevant scale, meaning that in general it is a direct sum of a topological sector and some further IR remnants, possibly of other kinds.

\medskip

In this section we assume that $\mathbf L_X$ is smooth and we focus on the topological sector corresponding to the torsional components of the defect group (and possible finite 0-form symmetries).\footnote{\ The case when $\mathbf{L}_X$ has singularities is more interesting to analyse. Typically, singularities are associated to gauge symmetries or more general interacting systems. The loci of said singularities can be interpreted as compactification manifolds for these systems, resulting in parameters. Since $\mathbf{L}_X$ has infinite volume, in the geometric engineering at infinity these parameters are set either to infinity or to zero. This limit has to be taken with care: see \cite{Bonetti:2024cjk} for a discussion of continuous non-abelian 0-form symmetries from this perspective.} In this case, as emphasised above, we expect the resulting bulk system to be a topological field theory of (generalised) Dijkgraaf-Witten type. We are in particular interested in the topological extended operators of $\mathcal F_X^{top}$ arising from this scaling limit. Conjecturally, the latter arise from wrapping the branes of the theory $\mathscr S$ on the torsional cycles of $\mathbf L_X$ at infinity \cite{GarciaEtxebarria:2022vzq,Apruzzi:2022rei, Heckman:2022muc, Bah:2023ymy, Apruzzi:2023uma,Etheredge:2023ler}. 
 
\medskip

\begin{figure}
    \centering
    \begin{tikzpicture}
        \draw[black] (-4,0,-1) -- (3.05,2,-0.3);
        \draw[black] (-4,0,-1) -- (2.75,-2,0.32);
        \draw[blue] (3,0,0.5)
            \foreach \t in {5,10,...,145}
                {--(3,{-0.5*sin(\t)},{0.5*cos(\t})}
            ;
        \draw[blue] (3, -0.17101, -0.469846)
        \foreach \t in {165,...,355}
                {--(3,{-0.5*sin(\t)},{0.5*cos(\t})}
        -- (3,0,0.5);
        
        \draw[blue] (-4,0,-1) -- (3.06,0.54,0);
        \draw[blue] (-4,0,-1) -- (2.94,-0.54,0);
        
        \draw[black] (3,0,2)
            \foreach \t in {5,10,...,355}
                {--(3,{-2*sin(\t)},{2*cos(\t})}
            -- (3,0,2);
        
        \draw[red] (3.1,0,0) 
            \foreach \t in {5,10,...,190}
                {--({3.1+0.25*sin(\t)},{-0.25+0.25*cos(\t)},{0.25*sin(\t)})}
                ;
        \draw[red] (2.95661, -0.454788, -0.143394)
            \foreach \t in {220,...,355}
                {--({3.1+0.25*sin(\t)},{-0.25+0.25*cos(\t)},{0.25*sin(\t)})}
            -- (3.1,0,0) ;
        \draw[purple, fill=purple!60] (-4,0,-1) circle (2pt);
        \node[left, purple] at (-4.1,0,-1){{\footnotesize SCFT}};
        \node[blue] at (0,0.6,-0.5){$\mathrm{Cone}(\beta)$};
        \node[blue] at (3.1,0.75,0){$\beta$};
        \node[red] at (3.25,-0.75,0){$\beta^\vee$};
        \node[black] at (3.4,2.4,0){$\mathbf{L}_X$};
        \node[black] at (1.4,2.0,0){$X$};

        \draw[black] (-10.5,-2,0) -- (-5.5,-2,0) -- (-5.5,2,0) -- (-10.5,2,0) -- (-10.5,-2,0);
        
        \draw[blue] (-7.5,0,0) 
            \foreach \t in {5,10,...,55}
                {--({-8.5+cos(\t)},{sin(\t)},0)}
            ;
        \draw[blue] (-8.07738, 0.906308)
            \foreach \t in {70,...,355}
                {--({-8.5+cos(\t)},{sin(\t)},0)}
        --(-7.5,0,0);
        \draw[red] (-8.5,0,0)
            \foreach \t in {5,10,...,55}
                {--({-7.5-cos(\t)},{-sin(\t)},0)}
            ;
            
        \draw[red] (-7.92262, -0.906308)
            \foreach \t in {70,...,355}
                {--({-7.5-cos(\t)},{-sin(\t)},0)}
        --(-8.5,0,0);
        \node[black] at (-10.6,2.3,0){$M$};
        \node[blue] at (-9.4,1.04,0){$\gamma$};
        \node[red] at (-6.5,1,0){$\gamma^\vee$};
    \end{tikzpicture}
    \caption{Schematic description of geometric engineering on a conical singular geometry. }
    \label{fig:cone}
\end{figure}
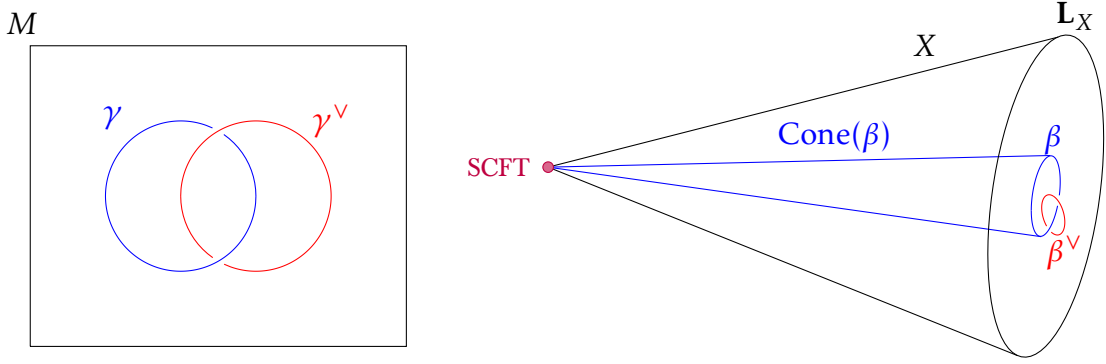

We denote the (extended) operators in $\mathcal F_X^{top}$ arising from membranes $\mathbf{Mp}$ wrapping torsional cycles in the boundary by \be \beta \in H_k(\mathbf{L}_X)\ee and supported in $D+1$ dimensions as \be\label{eq:topodefecto} \mathcal D_{\mathbf{Mp}}^\beta(\Sigma) \ee to distinguish them from the ones giving rise to defects $\mathcal W_{\mathbf{Mp}}^S$, that we have introduced in Equation \eqref{eq:wilson} in the previous section. In the geometric engineering at infinity, these membranes are topological membranes of $\mathcal F_X^{top}$, the extended topological operators of the theory. Here $\Sigma$ is a $(p-k+1)$-dimensional submanifold of the $(D+1)$-dimensional bulk of $\mathcal F_X^{top}$. These topological defects encode the symmetries of the relative theory $\mathcal T_X$. In particular, the charges of $\mathcal T_X$ are organised via the long exact sequence in relative homology
\be
\cdots \to H_{k+1}(X) \to H_{k+1}(X,\mathbf{L}_X) \to H_{k}(\mathbf{L}_X) \to H_{k}(X) \to \cdots.
\ee
In all the examples we consider in this paper, this long exact sequence truncates, which gives rise to interesting isomorphisms. Consider for example the case $H_{k}(X) = 0$ above. If this is the case, for a defect $\mathcal W_{\mathbf{Mp}}^S$ of $\mathcal T_X$ with $S \in H_{k+1}(X,\mathbf{L}_X) $ such that $[S] \neq 0$ in the defect group, there is a topological membrane in $\mathcal F^{top}_X$, namely $\mathcal D^{\beta}_{\mathbf{Mp}}$, obtained by wrapping a $k$-cycle $\beta \in H_k(\mathbf{L}_X)$ such that
\be
S = \text{Cone}(\beta).
\ee
This configuration is illustrated in blue in Figure \ref{fig:cone}. Its interpretation in field theory is that the extended operator $\mathcal W^S_{\mathbf{Mp}}(\gamma)$ inserted along an $n$-cycle $\gamma$ (where $n=p-k$) in a spacetime $M$, is a non-topological boundary condition for the topological defect $\mathcal D^\beta_{\mathbf{Mp}}(\Sigma)$ of $\mathcal F^{top}_X$, where $\Sigma$ is an $(n+1)$-cycle, exdended in the $D+1$ dimensional bulk and ending along the boundary $\{0\} \times M^D$ along $\gamma$.

\medskip
  
In many examples, the geometric engineering setup is such that:
\begin{enumerate}
\item There are membranes $\mathbf{Mp}$ and $\mathbf{Mp}^\vee$ in $\mathscr S$ which are electromagnetic dual;
\item There are torsional cycles in $\beta \in H_{k}(\mathbf L_X)$ and $\beta^\vee \in H_{k^\vee}(\mathbf L_X)$ that have a non-trivial linking pairing $\text{Link}_{\mathbf L_X}(\beta,\beta^\vee) \in \mathbb Q$, i.e. $k+k^\vee = $dim$(\mathbf L_X) - 1$;\footnote{\ Notice that because of the interplay between (higher-form) electromagnetic duality and Poincar\'e duality, the torsional parts of the groups $H_{k}(\mathbf L_X)$ and $H_{k^\vee}(\mathbf L_X)$ when non-trivial are necessarily isomorphic.}
\item The operators $\mathcal D_\mathbf{Mp}^\beta$ and $\mathcal D_\mathbf{Mp^\vee}^{\beta^\vee}$ are such that their supports are spheres linking in $D+1$ dimensions
\end{enumerate}
Then  $\mathcal F_X^{top}$ has topological operators of dimensions $n +1$ and $n^\vee + 1 = p^\vee - k^\vee + 1$ that satisfy
\begin{equation}\label{eq:heisenlink}
\langle \mathcal D_\mathbf{Mp}^\beta(\Sigma) \,\mathcal D_\mathbf{Mp^\vee}^{\beta^\vee}(\Sigma^\vee) \cdots \rangle \sim \text{exp}(2\pi \I \, \textrm{Link}_{\mathbf L_X}(\beta,\beta^\vee)\textrm{Link}_{D+1}(\Sigma,\Sigma^\vee)  ) \times \langle \cdots \rangle
\end{equation}
where $\textrm{Link}_{D+1}$ is the linking pairing of the $(n+1)$-cycle $\Sigma$ and the $n^\vee+1$ dimensional cycle $\Sigma^\vee$ in the $(D+1)$-dimensional space where $\mathcal F^{top}_X$ is defined. From this correlator, we see that pushing the $\mathcal D_\mathbf{Mp}^\beta$ and $\mathcal D_\mathbf{Mp^\vee}^{\beta^\vee}$ operators to the $D$-dimensional boundary $M$ where $\mathcal T_X$ is defined, one obtains a Heisenberg algebra acting on the Hilbert space as follows
\be
\mathcal D_\mathbf{Mp}^\beta(\Sigma) \,\mathcal D_\mathbf{Mp^\vee}^{\beta^\vee}(\Sigma^\vee) = \text{exp}\Big(2\pi \I \, \textrm{Link}_{\mathbf L_X}(\beta,\beta^\vee)(\Sigma \cap_{M}\Sigma^\vee)\Big)  \,\mathcal D_\mathbf{Mp^\vee}^{\beta^\vee}(\Sigma^\vee)\,\mathcal D_\mathbf{Mp}^\beta(\Sigma).
\ee
Whenever this phase is non-trivial, the Hilbert space $\mathcal F^{top}_X(M)$ is necessarily higher dimensional and the theory $\mathcal T_X$ is indeed relative. 

Moreover, considering the bounding cycle $\partial\Sigma = \gamma$ and ending the operator $\mathcal D^\beta_{\mathbf{Mp}}(\Sigma)$ along the defect $\mathcal W^S_{\mathbf{Mp}}(\gamma)$ as in the paragraph above we see that the $(D+1)$-dimensional correlator in equation \eqref{eq:heisenlink} pushes to the boundary as follows: 
\be\label{eq:actionofsymmetry}
\begin{aligned}
\mathcal T_X(  \mathcal D_{\mathbf{Mp}^\vee}^{\beta^\vee} (\Sigma^\vee) & \mathcal W_{\mathbf{Mp}}^S(\gamma) \cdots \subset M) \\
&= \text{exp} (2\pi \I \, \textrm{Link}_D(\gamma,\Sigma^\vee)\textrm{Link}_{\mathbf{L}_X}(\beta,\tilde{\beta})) \, \mathcal T_X( \mathcal W_{\mathbf{Mp}}^S(\gamma) \cdots \subset M),
\end{aligned}
\ee
hence, $\mathcal D_{\mathbf{Mp}^\vee}^{\beta^\vee} (\Sigma^\vee)$ has the interpretation of a generalised symmetry operator for the relative field theory $\mathcal T_X$, transforming uniformly its partition vector with charges dictated from the insertion of the defect $\mathcal W_{\mathbf{Mp}}^S(\gamma)$. In equation \eqref{eq:actionofsymmetry} we have indicated by $\cdots$ the presence of possible other insertions of operators and defects that are neutral with respect to the topological membrane $\mathcal D_{\mathbf{Mp}^\vee}^{\beta^\vee}$. The theory then has topological membranes of dimension $n^\vee + 1$, giving rise to an $n$-form generalised symmetry. Notice that, \textit{mutatis mutandis}, whenever $\mathbf{Mp}$ and $\textbf{Mp}^\vee$ in the above construction are different membranes, the relative theory also has an $n^\vee$-form generalised symmetry constructed in a similar way.\footnote{\ This is not the case for D3 branes in type IIB. Consider as an example the case of $\mathbb C^2/\mathbb Z_N$, which gives rise to a $D=6$ $(2,0)$ SCFT. In that context one has a defect group $\mathbb D^{(2)} = \mathbb Z_N$ that corresponds to charges of surface defects of the $(2,0)$ theory. The corresponding topological membranes arise from D3 branes wrapping the torsional $\mathbb Z_N$ cycles in $H_1(S^3/\mathbb Z_N)$, and indeed have supports of dimension 3. The latter link the two-dimensional defects, and because of the self duality give rise to a pairing of the form in \eqref{eq:actionofsymmetry}. For this reason the $D=6$ $(2,0)$ SCFTs are typically relative for generic values of $N$ --- see the detailed discussion in \cite{Gukov:2020btk}.} In this case the defect group of $\mathcal T_X$ has a subfactor
\be
\mathbb D^{(n)}_X \oplus \mathbb D^{(n^\vee)}_X,
\ee
where the two summands are mutually non-local. In the absence of higher linkings between the topological membranes of $\mathcal F^{top}_X$, the topological boundary conditions in presence of such an Heisenberg action on the Hilbert space are determined via the Stone-von Neumann theorem. The result under this assumption is that the topological boundary conditions are in one to one correspondence with the maximally isotropic sublattices $\Lambda$ of $\mathbb D^{(n)}_X \oplus \mathbb D^{(n^\vee)}_X$ with respect to the linking pairing. The resulting topological boundary conditions $\mathcal B_\Lambda$ set simultaneously all the operators with $\beta \in \Lambda$ to zero, and allow the definition of absolute theories $\mathcal T_{X}^{\mathcal B_\Lambda}$, with symmetry given by operators that remain non-trivial in the sandwitch construction. We omit a review of the details of this construction, and refer our readers to e.g. \cite{Freed:2006ya, Garcia-Etxebarria:2019caf, Morrison:2020ool, Albertini:2020mdx}. This analysis, however, needs to be refined in presence of non-trivial higher links along $\mathbf{L}_X$, because these can become obstructions to the existence of topological boundary conditions for the topological membranes of $\mathcal F^{top}_X$. In the next section we discuss how the higher linking obstructions arise in the geometric engineering dictionary.

\subsection{Anomalies, higher links, and geometry}\label{sec:higherlinks}

In the above example we have seen that a non-trivial linking pairing gives a geometrical origin to the obstruction of certain boundary condition to appear simultaneously. This is the counterpart of the non-commutativity of the fluxes associated with the branes engineering the defects and it was used to explain the emergence of different global structures of field theories from geometry \cite{Freed:2006ya, Garcia-Etxebarria:2019caf, Albertini:2020mdx}. A further profound aspect of the topological symmetry theory is that higher linking of topological operators provides further more refined information. Without prescribing such higher linking data, the information captured in the defect group is incomplete.

As an example, higher linking can be used to forbid the realisation of certain global forms, thus becoming obstructions to the existence of certain topological boundary conditions, or, equivalently, 't Hooft anomalies obstructing the gauging of the corresponding topological defects \cite{Kaidi:2023maf}.

In this section we propose a recipe to detect higher links of topological defects in $\mathcal F^{top}_X$ directly from geometry. In the following section we discuss some explicit examples. We refer our readers to Appendix \ref{app:torsionallinks} for the definitions of higher linking numbers we use in this Section.

The key equation is a straightforward generalisation of equation \eqref{eq:heisenlink}. If there are:
\begin{enumerate}
\item a collection of membranes $\mathbf{Mp}_a$, where $a=1,...,L$, of the string theory $\mathscr S$ that correspond to topological membranes dual to torsional charges in the defect group $\mathbb D_X$;\footnote{\ Note that for given $a$ the membrane $\mathbf{Mp}_a$ has dimension $p_a + 1$ in our conventions.}
\item a Brunnian configuration of torsional cycles in homology $\beta^a \in H_{k_a}(\mathbf{L}_X)$ such that these have a non-trivial higher linking pairing
\be
\text{Link}^{(\varkappa)}_{\mathbf L_X}(\beta^1,...,\beta^L) \in \mathbb Q\,;
\ee
\item a Brunnian configuration of cycles $\Sigma^a$ with $a=1,...,L$ with dimensions 
\be
\dim \Sigma^a = p_a - k_a + 1\,,
\ee
that are topological spheres such that they form a non-trivial higher link in $S^{D+1}$ 
\be
\mathrm{Link}^{(\kappa)}_{S^{D+1}}(\Sigma^1,...,\Sigma^L) \in \mathbb Z\,.
\ee
Notice that the latter is always integer valued because the sphere has no torsion;
\end{enumerate}
then, we claim that the following correlator is non-trivial
\begin{equation}
\label{eq:branelink}
\begin{aligned}
&\left\langle \prod_{a=1}^{L} \mathcal D_{\mathbf{Mp}_a}^{\beta^a}(\Sigma^a) \cdots \right\rangle_{{\mathcal F^{top}_X}(S^{D+1})} =\\
&\qquad\qquad=\mathrm{exp}\left(2\pi \I \, \text{Link}^{(\varkappa)}_{\mathbf L_X}(\beta^1,\dots, \beta^L)\mathrm{Link}^{(\kappa)}_{S^{D+1}}(\Sigma^1,\dots,\Sigma^L)\right) \langle\cdots \rangle_{{\mathcal F^{top}_X}(S^{D+1})} 
\end{aligned}
\end{equation}
in the SymTFT. Here $\cdots$ denotes the presence of other operators that are not participating in the link (and/or are neutral with respect to the corresponding topological membranes). Notice that in \cref{eq:branelink}, we are considering all possible links among the topological membranes that are dual to charges in the defect group, not just the ones that are electro-magnetic dual pairs.

\medskip

We stress that while when $L = 2$ above, the structure of the linking paring is essentially unique, whenever there are more than two components the possible linking pairings become much richer. Moreover, there are two different linkings in the formula \eqref{eq:branelink}, that correspond to the fact that while along $\mathbf L_X$ we are linking $L$ distinct \textit{torsional} $k_a$-cycles, along the $(D+1)$-dimensional spacetime where $\mathcal F^{top}_X$ is defined we are linking non-torsional cycles. The linking for torsional cycles along the boundary is slightly more complicated to construct due to the lack of Seifert surfaces -- however for the examples we consider in this paper, it is always possible to do so (we refer our readers to Appendix \ref{app:torsionallinks} for these technical details). In summary
 
\begin{itemize}
\item A necessary condition for forming a link (of type $\kappa$) between $L$ distinct cycles $\Sigma^a$ in $D+1$ dimensions (here $a=1,...,L$)  is that the degree of the various cycles involved satisfy
\begin{align}\label{eq:linkm}
    \sum_{a=1}^L \dim \Sigma^a + L-\kappa = (L - 1) (D+1) \, ,
\end{align}
where $ 0 \leq \kappa < L -1$;
\item A necessary condition for forming a link (of type $\varkappa$) among the $L$ torsional cycles $\beta^1,...,\beta^L$ in $\mathbf L_X$ is that their degrees $k_a$ satisfy
\begin{align}\label{eq:highertorcia}
   \sum_{a=1}^L k_a +L-\varkappa = (L-1) \dim \mathbf{L}_X.
\end{align}
where $0 \leq \varkappa < L -1$.
\end{itemize}
The former type of link gives rise to a non-trivial contribution only if the second type of link gives a non-trivial phase in equation \eqref{eq:branelink}. In the following sections we will give explicit examples of how to compute these quantities from geometry. It is interesting to remark that the higher links with $L$ strands of type $\varkappa = L-1$ seem to play a special role --- see the discussion in Appendix \ref{app:torsionallinks}.

\paragraph{General structure of links from branes.} Before diving into a detailed discussion of specific examples, let us remark that for a background of the form
\be
X_1 \times X_2 \times ... \times X_K,
\ee
where $\sum_{i=1}^K \dim X_i = D_{\mathscr S}$, one can generalise the formulae above and consider collections of $L$ strands of branes that form $L$-links of different kinds over the various components of spacetime. This allows us to relax the constraint on dimensions as follows,
\be
\sum_{p \text{ branes}} N_{p} (p + 1) + K L  - \sum_{j=1}^K \kappa_j = (L -1)D_{\mathscr S}
\ee
where we are considering a link with 
\be 
L = \sum_{p \text{ branes}} N_p
\ee
many strands, out of which $N_p$ are formed out of $p$-branes. For example, let us consider the M-theory case. We take the geometric engineering limit at infinity of M-theory on a background of the form
\be
S^{D+1} \times L_X
\ee
where $\dim L_X = d-1$ and $D+d = 11$. Then we have the following constraint:
\be
5 N_2 + 8 N_5 - \kappa_{S^{D+1}} - \varkappa_{L_X} = (N_2 + N_5 - 1) 11,
\ee
where $\kappa_{S^{D+1}}$ and $\varkappa_{L_X}$ can both vary between $1$ and $L$. We notice that by choosing $\varkappa_{L_X} = N_2 + N_5 -1$, we obtain 
\be
7 N_2 + 4 N_5 = 12 - \kappa_{S^{D+1}}.
\ee
From the above constraint it is easy to see that only $\kappa_{S^{D+1}} =0,1$ links are permitted, which give rise to the links obtained so far from dimensional reduction of higher Chern-Simons terms in suitable truncations of stringy effective actions \cite{Apruzzi:2021nmk, GarciaEtxebarria:2019caf, Apruzzi:2023uma}.

\medskip

Before we go to examples, however, let us pause to present a remark abount the interplay with the boundary. This can occur, whenever some of the cycles $\Sigma^a$ are such that $\partial \Sigma^a = \gamma^a$ for a cycle $\gamma^a$ along the boundary, and we have a corresponding boundary defect $\mathcal W^{S^a}_{\mathbf{Mp}_a}(\gamma^a)$ in $\mathcal T_X$ where the topological operator $\mathcal D^{\beta^a}_{\mathbf{Mp}_a}(\Sigma^a)$ ends (and so in particular, we have that $S^a = \text{Cone}(\beta^a)$). In this case, a non-trivial link of the type in \eqref{eq:branelink} can give rise to higher actions on collections of intersecting and linking defects, just by pushing the correlator in $\mathcal F^{top}_X$ onto the boundary $\mathcal T_X$. It is interesting, but beyond the scope of this note, to investigate these effects in detail and their repercussions on the higher structure of symmetry. 

\bigskip

At present, we will limit ourselves to exploit our formulae to recover some known results about duality defects in 4D $\mathcal N=4$ SYM, as well as several known results in the case of M-theory, that have been obtained via the hybrid formalism \cite{Apruzzi:2021nmk, Apruzzi:2023uma, vanBeest:2022fss, Lawrie:2023tdz, Heckman:2022muc}. This serves as a first collection of non-trivial consistency checks . The analysis in this section can be refined and extended whenever $\mathbf L_X$ has orbifold singularities --- which often signals the presence of continuous 0-form symmetries and is ubiquitous in the context of geometric engineering (see e.g. \cite{Acharya:2023bth} for a recent discussion).

\section{Examples}\label{sec:examples}

In this section we give some concrete examples of applications of the dictionary we discussed above --- other interesting applications and generalisations are discussed in  \cite{DMDZM,BDZM1}. In section \ref{sec:7d} we quickly review the case of 7D SYM from M-theory from our perspective. In section \ref{sec:N=4revisited} we discuss the non-invertible duality defects of 4D $\mathcal N=4$ $\mathfrak{su}(N)$ theories, thus recovering the previous results by \cite{Kaidi:2021xfk,Choi:2021kmx} in a geometric engineering framework. In \ref{sec:5d} we reproduce within our formalism an interesting obstruction to the existence of a magnetic phase in 5D theories with an $\SU(p)_q$ gauge theory phase \cite{Apruzzi:2021nmk, Gukov:2020btk, BenettiGenolini:2020doj}.

\subsection{Seven dimensional SYM theories}\label{sec:7d}

As a warm-up example, let us consider M-theory on $X=\mathbb C^2/ \Gamma$ where $\Gamma$ is a finite subgroup of $\SU(2)$. These spaces all have conical ALE metrics. It is well known that M-theory on such an orbifold background engineers seven dimensional super Yang-Mills (SYM) with a gauge group whose corresponding Lie algebra $\mathfrak{g}$ is of $ADE$ type (via the McKay correspondence) --- see e.g. \cite{Sen:1997js} or \cite{Tachikawa:2015wka} for a review. 

\medskip

In the $D=7$ theory, the extended defects of the theory, i.e. Wilson lines and 't Hooft 4-dimensional magnetic defects, are engineered wrapping M2 branes and M5 branes on non-compact 2-cycles, respectively. The defect group can be easily computed from  the boundary geometry, which in this case is $\mathbf{L}_X= S^3/ \Gamma$. The homology of $\mathbf{L}_X$ is then given by
\be
H_\bullet (S^3/ \Gamma) = \{\mathbb Z, \text{Ab}(\Gamma), 0 ,\mathbb Z\}\,,
\ee
with the only non-trivial torsional part being $H_1 (S^3/ \Gamma)$.  The formula in \cref{eq:defectgroup} boils down to 
\begin{align}
\text{Tor } \mathbb{D}_X = \text{Ab}(\Gamma)^{(1)}_{M2} \oplus \text{Ab}(\Gamma)^{(4)}_{M5} .
\end{align}
In this context, the dual topological membranes arise from wrapping M2s and M5s on the corresponding torsional cycles, labeled by $\text{Ab}(\Gamma)$. The resulting topological membranes are:
\be
\mathcal D^{\beta}_{M2}(\Sigma^2), \qquad 
\mathcal D^{\beta}_{M5}(\Sigma^5)  .
\ee
Clearly, $\mathcal D^{\beta}_{M2}(\Sigma^2)$ can end along a Wilson line $\mathcal W^{\text{Cone}(\beta)}_{M2}(\gamma^1)$ and $\mathcal D^{\beta}_{M5}(\Sigma^5)$ can end along a 't Hooft hypersurface $\mathcal W^{\text{Cone}(\beta)}_{M5}(\gamma^4)$. These topological membranes have a non-trivial linking summarised in Table \ref{ADE-linking}. As expected, their linking measures the charge of the given line with respect to the center of the gauge group. 

\medskip

\begin{table}
$$  
  \arraycolsep=4.4pt\def\arraystretch{1.2}
  \begin{array}{cccc}
    \Gamma & \mathfrak g_\Gamma & \text{Ab}(\Gamma) &  \ell_{\mathbf{L}_X} \\
    \hline
    \mathbb Z_N & \mathfrak{su}(N) & \mathbb Z_N & \frac{1}{N} \\
    \mathrm{Dic}_{(4N-2)} & \mathfrak{so}(8N) & \mathbb Z_2\oplus \mathbb Z_2 & \left(\begin{matrix}0 & 1/2 \\ 1/2 & 0\end{matrix}\right) \\
    \mathrm{Dic}_{(4N-1)} & \mathfrak{so}(8N+2) & \mathbb Z_4 & \frac{3}{4} \\
    \mathrm{Dic}_{(4N)} & \mathfrak{so}(8N+4) & \mathbb Z_2\oplus \mathbb Z_2 & \left(\begin{matrix}1/2 & 0 \\ 0 & 1/2\end{matrix}\right)\\
    \mathrm{Dic}_{(4N+1)} & \mathfrak{so}(8N+6) & \mathbb Z_4 & \frac{1}{4} \\
    2T & \mathfrak{e}_6 & \mathbb Z_3 & \frac{2}{3} \\
    2O & \mathfrak{e}_7 & \mathbb Z_2 & \frac{1}{2} \\
    2I & \mathfrak{e}_8 & 0 & 0
  \end{array} 
$$
\caption{Linking pairings for $S^3/\Gamma$ from \cite{GarciaEtxebarria:2019caf}.}\label{ADE-linking}
\end{table}

In this context, all $k_a = 1$ and a straightforward application of the formula in equation \eqref{eq:highertorcia} shows that, on top of the case $L=2$, $\varkappa=1$, giving rise to the Heisenberg algebra of non-commuting fluxes, there is another possibly non-trivial linking pairing on $\mathbf{L}_X$, namely for $L=3$ and $\varkappa =0$. In this latter case, however, one finds no solution to \eqref{eq:linkm} for a link with three strands that are either two or five dimensional in 8 dimensions, where the relevant SymTFT is supported. Hence, there are no higher linking terms for the topological membranes in the resulting topological theory $\mathcal F^{top}_X$. We therefore conclude that the possible boundary conditions that have been classified in \cite{Garcia-Etxebarria:2019caf, Albertini:2020mdx} are all unobstructed.\footnote{\ Consequences of the discrete 0-form symmetries which arise from the isometries in these examples are discussed in greater detail in \cite{DMDZM}, where also the corresponding dimensional reductions are characterised in details.}

\subsection{Duality defects in \texorpdfstring{$\mathcal N=4$ $\mathfrak{su}_N$}{N=4 su(N)} SYM from IIA}\label{sec:N=4revisited}

 It is well-known that one way to obtain $\mathcal N=4$ $\mathfrak{su}_N$ SYM theories geometrically is to consider the geometric engineering limit of the IIA superstrings on the space 
\be\label{eq:IIAGEO}
X = T^2 \times \mathbb C^2/\mathbb Z_N.
\ee
Here the complexified gauge coupling of the theory is identified with the complexified volume of the torus, that we denote 
\be
\tau = \frac{1}{2\pi}\int_{T^2} B_{NS} + \I R_1 R_2
\ee
where $R_1$ and $R_2$ are the radii of the two circles in suitable units. The W-bosons of the theory arise from D2 branes wrapped on the vanishing 2-cycles of the ALE singularity. Recall that resolving the ALE singularity corresponds to blowing up a collection $N-1$ 2-spheres, that we denote $\mathbb P^1_a$ ($a=1,...,N-1$). To each such 2-sphere corresponds a harmonic 2-form $\omega_a$, that gives a total of $(N-1)$ many Maxwell fields by decomposing the RR $C_3$ potential. These fields correspond to the Cartan subalgebra of the gauge group. The other W-bosons arise from stable BPS boundstates of D2 branes wrapping collections of these $\mathbb P^1$ which are in one-to-one correspondence with the positive roots of $\mathfrak{su}_N$ \cite{Witten:1995ex,Sen:1997js}. Furthermore, D4 branes wrapped on $T^2 \times \mathbb P^1_a$ correspond to magnetic monopoles. One can also consider wrapped D2-D4 boundstates, which then correspond to dyons.

In this context $S$-duality is realised via T-duality on the two circles in the torus. The latter corresponds to the self-duality of IIA upon volume inversion of the $T^2$ which gives the stringy origin of Montonen-Olive duality \cite{Montonen:1977sn,Vafa:1997mh}. Upon these two T-dualities the D2 branes wrapped on $\mathbb P^1_a$ become D4 branes wrapped on $T^2_{\text{dual}} \times \mathbb P^1_a$. Conversely, the D4 branes wrapped on $T^2 \times \mathbb P^1_a$ are mapped to D2 branes wrapped on $\mathbb P^1_a$ in the dual geometry. This exchanges electric and magnetic degrees of freedom. Consider now a torus fixed at the self-dual value of $\tau = \I$, which is obtained by choosing $R_1 = R_2 = 1$ and setting $\int_{T^2} B_{NS}$ to zero. Let us denote it $T^2_*$. The geometry $T^2_* \times \mathbb C^2/\mathbb Z_N$ acquires an extra (quantum) $(\mathbb Z_2)^{(0)}_{T^2}$ symmetry corresponding to the fact that this geometry is invariant upon the volume inversion in IIA. Let us denote the corresponding codimension 1 defect $\mathcal S^{(0)}$.

Now we consider the extended operators of this theory. The space $X$ has a non-trivial boundary
\be
\partial X = T^2 \times S^3/\mathbb Z_N,
\ee
whose homology is given by
\be\label{eq:notN4}
H_\bullet(T^2 \times S^3/\mathbb Z_N) = \big\{\mathbb Z,\, \mathbb Z^2 \oplus \mathbb Z_N, \, \mathbb Z \oplus \mathbb Z_N^2, \, \mathbb Z \oplus \mathbb Z_N, \, \mathbb Z ^2,\, \mathbb Z  \big\}.
\ee
Exploiting the conical metric on the ALE singularity $\mathbb C^2/\mathbb Z_N$ we can give a geometrical origin for the defects of this theory. Focusing on the torsional cycles, we see that the theory has Wilson lines arising from D2 branes wrapped along the non-compact 2-cycles 
\be
\mathcal W^{\text{Cone}_{ALE}(\alpha)}_{D2}
\ee
for all $\alpha \in \mathrm{Tor}\, H_1(T^2 \times S^3/\mathbb Z_N)$, as well as 't Hooft lines arising from D4 branes wrapping the non-compact 4-cycles 
\be
\mathcal W^{\text{Cone}_{ALE}(\alpha) \times T^2}_{D4}
\ee
corresponding to the cycles $\beta \sim \alpha \times T^2  \in \mathrm{Tor}\, H_3(T^2 \times S^3/\mathbb Z_N)$. The resulting theory then has a corresponding 1-form defect group
\be
\mathbb D^{(1)} = (\mathbb Z_N)_{D2} \oplus (\mathbb Z_N)_{D4}
\ee
corresponding to the charges of these defects. Since D2 and D4 branes are not mutually local, the theory $\mathcal T_{X} = GE_{IIA/X}$ is indeed a relative field theory. The corresponding boundary theory has topological membranes
\be
\mathcal D^\alpha_{D2}(\Sigma^2) \qquad \mathcal D^\beta_{D4}(\Sigma^2).
\ee
The $\mathcal D^\alpha_{D2}(\Sigma^2)$ membranes can end along $\mathcal W^{\text{Cone}_{ALE}(\alpha)}_{D2}$ lines on the boundary and similarly the  $\mathcal D^\beta_{D4}(\Sigma^2)$ topological membranes can end along the $\mathcal W^{\text{Cone}_{ALE}(\alpha) \times T^2}_{D4}$ 't Hooft lines. The corresponding linking pairing is induced from the one along torsional cycles along the ALE boundary (see Table \ref{ADE-linking} above), and one indeed obtains that
\be
\mathcal D^\alpha_{D2}(\Sigma^2)\mathcal D^\beta_{D4}(\widetilde{\Sigma}^2) \sim \text{exp}\left(2 \pi \I \, \frac{\alpha \beta}{N} \textrm{Link}_5(\Sigma^2,\widetilde{\Sigma}^2)  \right).
\ee
which by pushing it to the boundary as discussed above induced the expected action of topological symmetries on Wilson and 't Hooft operators for 4D SYM. There are no higher linking constraining further the algebra of the topological defects contributing to the 4D $\mathcal N=4$ spectrum --- we give a detailed analysis in Appendix \ref{app:KK4DN=4}.\footnote{\ As discussed in more detail in Appendix \ref{app:KK4DN=4} the spectrum of defects one obtains from Equation \eqref{eq:notN4} correspond to the toroidal reduction of the defect group of a 6d (1,1) $\mathfrak{su}_N$ gauge theory. This is indeed consistent: the theory $GE_{IIA/(\mathbb C^2/\mathbb Z_N)}$ is 6d (1,1) $\mathfrak{su}_N$, and the geometry we are considering is a toroidal reduction of the latter. Of course decoupling all the KK modes one lands to the relative 4D $\mathcal N=4$ $\mathfrak{su}_N$ SYM theory.}

Notice that at the self-dual radius, $\mathcal S^{(0)}$ generates a $(\mathbb Z_2)^{(0)}_\mathcal{S}$ symmetry of the relative theory $\mathcal T_{X} = GE_{IIA/X}$, since the self-duality of IIA has the effect of swapping the two factors of $\mathbb D^{(1)}$. This is just electromagnetic duality for the 5D topological theory $\mathcal F^{top}_X$. The possible gapped boundary conditions in this context are well-known to be in one-to one correspondence with the possible choices of maximally isotropic sublattices of $\mathbb Z_N \times \mathbb Z_N$. A well-known consequence of the Stone-von Neumann theorem is that all these boundary conditions are connected by discrete gauging operations (for a review see e.g. \cite{Witten:2009at}). In this context there is no higher linking which obstructs these topological boundaries, an easy consequence of \eqref{eq:highertorcia}. In this way we have geometrised all the ingredients necessary to reproduce geometrically the analysis by Kaidi, Zafrir and Zheng \cite{Kaidi:2022uux}.

For instance, choosing a boundary condition $\mathcal B_{D4}$ which sets to zero all the D4 charges at the boundary, realises the $\SU(N)$ $\mathcal N=4$ with a trivial background for its $(\mathbb Z_N)^{(1)}_{D2}$ electric 1-form symmetry. Implementing $\mathcal S^{(0)}$ maps this boundary condition to $\mathcal B_{D2}$ which gives the $\PSU(N)$ $\mathcal N=4$ theory with a trivial background for its $(\mathbb Z_N)^{(1)}_{D4}$ magnetic 1-form symmetry. However, by performing a discete gauging of this subgroup of the 1-form defect group of this theory on half of the space time gives another topological interface $\sigma^{0}$ that maps $\PSU(N)$ back to $\SU(N)$. Composing these two interfaces we obtain a new one $\mathcal N^{(0)} = \mathcal S^{(0)} \sigma^{(0)}$ which is a non-invertible duality defect for $\SU(N)$ \cite{Kaidi:2022uux}.

\medskip

There is of course a wide variety of geometrical interfaces arising from this effect, that give the geometrical counterpart of the constructions of \cite{Choi:2021kmx,Kaidi:2021xfk,Bashmakov:2022jtl,Bashmakov:2022uek} and also offer opportunities to wider generalisation, obtained combining fiberwise self-T-dualities of IIA and IIB superstrings with various geometric engineering backgrounds that are $T^2$ fibrations. Similarly, one can exploit the self-duality of M-theory upon $T^3$ volume inversion. Studying these effects would take us too far from the modest scope of this short note, and we defer the discussion of these more general duality defects from geometric engineering to our future work \cite{DMDZM}.

\subsection{\texorpdfstring{$\SU(p)_q$}{SU(p)q} 5D SCFTs}\label{sec:5d}

As a final class of examples in this paper we consider 5D SCFTs with $\SU(p)_q$ gauge theory phases. We choose these models because in \cite{Gukov:2020btk} and \cite{BenettiGenolini:2020doj}, using field theoretical methods, the authors were able to compute 't Hooft anomalies for the electric 1-form symmetry and mixed anomalies between 1-form and 0-form instanton symmetry. The result was recovered in \cite{Apruzzi:2021nmk} exploiting the dimensional reduction of the topological term of M-theory. The obstruction to gauging the 1-form symmetry is particularly interesting since it obstructs candidate global structures. In this section we are interested in recovering it from the perspective of topological branes and their higher linking at infinity.

\subsubsection{Geometric perspective}

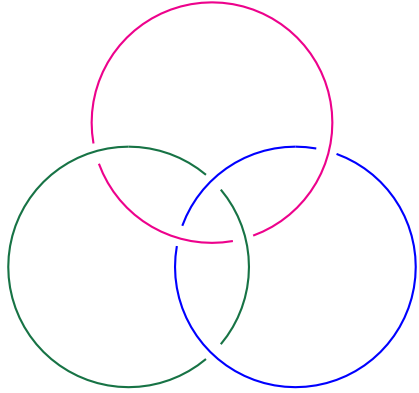
\begin{figure}
    \centering
    \begin{tikzpicture}[scale=1.6]
        \draw[magenta,thick] (0,1.8) 
            \foreach \t in {5,10,...,160}
                {--({sin(\t)},{0.8+cos(\t)})};
    
        \draw[magenta,thick] ({sin(170)},{0.8+cos(170)}) 
            \foreach \t in {175,180,...,250}
                {--({sin(\t)},{0.8+cos(\t)})};    

        \draw[magenta,thick] ({sin(260)},{0.8+cos(260)}) 
            \foreach \t in {265,270,...,355}
                {--({sin(\t)},{0.8+cos(\t)})} 
            -- (0,1.8);

        \draw[blue,thick] ({0.8*sin(120)},{0.8*cos(120)+1}) 
            \foreach \t in {5,10}
                {--({0.8*sin(120)+sin(\t)},{0.8*cos(120)+cos(\t)})};

        \draw[blue,thick] ({0.8*sin(120)+sin(20)},{0.8*cos(120)+cos(20)}) 
            \foreach \t in {25,30,...,280}
                {--({0.8*sin(120)+sin(\t)},{0.8*cos(120)+cos(\t)})};

        \draw[blue,thick] ({0.8*sin(120)+sin(290)},{0.8*cos(120)+cos(290)}) 
            \foreach \t in {295,300,...,355}
                {--({0.8*sin(120)+sin(\t)},{0.8*cos(120)+cos(\t)})}
            -- ({0.8*sin(120)},{0.8*cos(120)+1}) ;

        \draw[dgreen, thick] ({0.8*sin(240)},{0.8*cos(240)+1}) 
            \foreach \t in {5,10,...,40}
                {--({0.8*sin(240)+sin(\t)},{0.8*cos(240)+cos(\t)})};

        \draw[dgreen, thick] ({0.8*sin(240)+sin(50)},{0.8*cos(240)+cos(50)}) 
            \foreach \t in {50,55,...,130}
                {--({0.8*sin(240)+sin(\t)},{0.8*cos(240)+cos(\t)})};

        \draw[dgreen, thick] ({0.8*sin(240)+sin(140)},{0.8*cos(240)+cos(140)}) 
            \foreach \t in {140,145,...,355}
                {--({0.8*sin(240)+sin(\t)},{0.8*cos(240)+cos(\t)})}
            -- ({0.8*sin(240)},{0.8*cos(240)+1}) ;
    \end{tikzpicture}
    \caption{A Borromean link. Such a link is an example of a Brunnian link; that is, a non-trivial link that becomes isotopic to the unlink upon removal of any single connected component.}\label{fig:borro}
\end{figure}

The Calabi-Yau singularities $X$ corresponding to the SCFTs of interest in this section are metric cones over the Sasaki-Einstein $Y^{p,q}$ manifolds constructed in \cite{Martelli:2004wu}. For a detailed review of these geometries see e.g. \cite{Closset:2018bjz}. Here, we summarise the features we need for our analysis. These toric CYs have a toric fan with four vertices
\be
w_0 = (0,0), \qquad w_p = (0,p), \qquad w_{x} = (-1,q_x), \qquad w_{y} = (1,q_y),
\ee
such that
\be
q = p - (q_x + q_y).
\ee
The compact divisors corresponding to the generators of the Cartan torus corresponds to the point in the toric fan $S_a \leftrightarrow (0,a)$ for $a=1,...,p-1$. As discussed e.g. in Section 4.2 of \cite{Albertini:2020mdx}, the generator of the center symmetry corresponds to the divisor
\be
Z = \sum_{a=1}^{p-1} a S_a,
\ee
so that the charge of a given BPS particle obtained by wrapping a curve $\mathcal C$ with respect to the center is given by
\be
Q(\mathcal C) = Z \cdot \mathcal C\ \ (\mathrm{mod}\, p).
\ee
We want to recover the algebra of operators for the topological symmetry theory from the link of this singularity. These spaces have torsion in homology 
\be
\text{Tor}\,H_1(Y^{p,q}) \simeq \text{Tor}\,H_3(Y^{p,q}) \simeq \mathbb Z_{\gcd(p,q)}
\ee
The corresponding topological operators are
\be
\mathcal D^{\beta^1}_{\mathbf{M2}}(\Sigma^2), \qquad\qquad \mathcal D^{\beta^3}_{\mathbf{M5}}(\Sigma^3),
\ee
where $\beta^i$ is the torsional generating cycle for each factor. These operators have a non-trivial linking
\be\label{eq:Heisenbergo5d}
\langle \mathcal D^{\beta^1}_{\mathbf{M2}}(\Sigma^2) \mathcal D^{\beta^3}_{\mathbf{M5}}(\Sigma^3) \rangle \sim \text{exp}\left(2\pi \I \, \mathrm{Link}_6(\Sigma^2,\Sigma^3) \, \textrm{Link}_{Y^{p,q}}(\beta^1,\beta^3)\right),
\ee
where (see equation (B.7) of \cite{Albertini:2020mdx} for a derivation)
\be
\textrm{Link}_{Y^{p,q}}(\beta^1,\beta^3) = \frac{1 }{\text{gcd}(p,q)}.
\ee
This is not the only (higher) linking that we can form. As discussed in Appendix \ref{app:Ypq}, the necessary conditions for a higher linking are met also for the configuration
\be
\begin{aligned}
\langle  \mathcal D^{\beta^3}_{\mathbf{M5}}(\Sigma^3)_1  \mathcal D^{\beta^3}_{\mathbf{M5}}(\Sigma^3)_2  &\mathcal D^{\beta^3}_{\mathbf{M5}}(\Sigma^3)_3 \rangle \\
&\sim \text{exp}\left(2\pi \I \, \mathrm{Link}_6(\Sigma^3_1,\Sigma^3_2,\Sigma^3_3) \, \textrm{Link}^{(2)}_{Y^{p,q}}(\beta^3,\beta^3,\beta^3)\right).
\end{aligned}
\ee
Choosing $\Sigma^3_i =  S^3$ arranged as the Borromean link in Figure \ref{fig:borro}, the triple higher linking evaluates to one. We are left to compute $\textrm{Link}^{(2)}_{Y^{p,q}}(\beta^3,\beta^3,\beta^3)$. Since $\beta^3$ is torsional it is such that $\gcd(p,q) \beta^3 = \partial \gamma^4$ for a 4-cycle $\gamma^4$, that serves as a Seifert surface for $\gcd(p,q) \beta^3$ in this case. Starting from the triple linking with $\varkappa = 2$ we have (from Equation \eqref{eq:torsionhigherlink})
\be
\begin{aligned}
\textrm{Link}^{(2)}_{Y^{p,q}}(\beta^3,\beta^3,\beta^3)&=\left({1\over  \gcd(p,q)}\right)^3 \int_{Y^{p,q}} \mathrm{PD}(\gamma^4) \wedge \mathrm{d PD}(\gamma^4) \wedge \mathrm{d PD}(\gamma^4) \\
&= \left({1\over  \gcd(p,q)}\right)^3 \int_{X} \mathrm{PD}(Z) \wedge \mathrm{PD}(Z) \wedge \mathrm{PD}(Z)\\
&= \left({1\over  \gcd(p,q)}\right)^3 \times (Z \cdot Z \cdot Z)\\
&= {p (p - 1) (p^2 + p q - 2 q)\over  (\gcd(p,q))^3}\\
& = { q p (p - 1) (p-2) \over  (\gcd(p,q))^3 } +  \underbrace{{p^3(p-1) \over  \gcd(p,q)^3}}_{\in\, \mathbb Z}.
\end{aligned}
\ee
The triple intersection of $Z$ was determined in section 4.3 of \cite{Apruzzi:2021nmk}, and we have used Stokes' theorem with $\partial X = Y^{p,q}$, identitfying $d\mathrm{PD}(\gamma^4)$ in $Y^{p,q}$ with $\mathrm{PD}(Z)$ in $X$. As a result we obtain that
\be\label{eq:triplegeomen}
\langle  \mathcal D^{\beta^3}_{\mathbf{M5}}( S^3)_1  \mathcal D^{\beta^3}_{\mathbf{M5}}( S^3)_2  \mathcal D^{\beta^3}_{\mathbf{M5}}( S^3)_3 \rangle \sim \text{exp}\left(2\pi \I \, { q p (p - 1) (p-2) \over  (\gcd(p,q))^3 }  \right).
\ee
for a configuration of 3 three-spheres arranged according to a Borromean 3-link in 6 dimensions. This is the hallmark of an obstruction to gauging the topological symmetry generated by the operators $\mathcal D^{\beta^3}_{\mathbf{M5}}( S^3)$. This is an example of how the geometry of the singularity link enriches the structure of the defect group. To complete this discussion, in Section \ref{sec:SymTFTpers} we reproduce this result using the purely field theoretical perspective of \cite{Gukov:2020btk}. It is interesting to remark that the SymTFT analysis presented there implies that in presence of such higher linking, the topological operators $\mathcal D^{\beta^3}_{\mathbf{M5}}(S^3)$ become non-invertible.

\subsubsection{SymTFT perspective}\label{sec:SymTFTpers}

\paragraph{Anomalies from field theory.}\label{sec:ex5d}
Consider the $\SU(p)_q$ theory in five dimensions: we want to use the methodology advocated in \cite{Gukov:2020btk} and \cite{BenettiGenolini:2020doj} to recover the geometric result in section \ref{sec:5d}. 

Recall that the Lagrangian for the $\SU(p)_q$ theory is given by
\begin{align}\label{5dsymcs}
    \mathcal{L}= \frac{1}{g^2} \text{tr}( F \wedge * F )+ \frac{q}{6 (2 \pi)^2} \text{ tr} ( A \wedge F \wedge F).
\end{align}
Because of the Chern-Simons (CS) term, the 1-form symmetry acting on Wilson loops is broken to $\mathbb{Z}_{n}$ where $n = \gcd(p,q)$. As discussed in \cite{Gukov:2020btk}, one can can extend the CS term to a 6d manifold and extend the $\SU(p)$ connection to a $\U(p)$ connection with the constraint that $\mathrm{tr}(F_{\U(p)}-B_2 1_p)=0$, where $B_2$ is a background for the electric 1-form symmetry\footnote{\ As discussed in \cite{Gaiotto:2017yup}, this is the standard procedure to couple a 1-form symmetry to a background.}. We have then
\begin{align}\label{anomaly5d}
\begin{split}
    \int_{M_6} \diff(\mathrm{CS})[B_2] &=2 \pi \frac{q}{6 (2 \pi)^3}\int_{M_6}  \,\mathrm{tr} ( (F-B_2) \wedge (F-B_2) \wedge (F-B_2))  
    \\
    &= 2 \pi \frac{q p (p-1) (p-2)}{6 (2 \pi)^3} \int_{M_6} B_2 \wedge B_2 \wedge B_2 \quad  (\mathrm{mod}\,2\pi).
    \end{split}
\end{align}
This represents a 't Hooft anomaly for the 1-form symmetry. Since $B_2$ is a background for the $\mathbb{Z}_{n}$ 1-form, we have that
\begin{align}
    n B_2 = \diff \lambda_1 ,
\end{align}
thus the 't Hoof anomaly is trivial on spin manifolds, i.e. an integer modulo $2 \pi$, if
\begin{align}\label{anomaly-condition}
    \frac{q p (p-1) (p-2)}{n^3} \in \mathbb{Z} \, .
\end{align}
There are many implication to this statement. Firstly, for general $p$ and $q$, one cannot gauge the electric 1-form symmetry to go to the magnetic phase of the theory, i.e. to a wannabe $\mathrm{PSU}(p)_q$ theory. In the next section we will reinterpret this result in term of the presence of non trivial topological linking of the operators in the symmetry theory. Secondly, for specific values of $p$ and $q$ one can still gauge a subgroup of the 1-form symmetry. For example, $\SU(4)_{4}$, i.e. $q=4$, has a $\mathbb{Z}_4$ 1-form symmetry which is anomalous, but has a $\mathbb{Z}_2$ subgroup can be gauged giving a 5D $\mathrm{SO}(6)$ theory \cite{Gukov:2020btk}.

In the following we will revisit this anomaly in term of the data in the $\mathcal{F}^{top}_X$ theory, i.e. the twisted $BF$-theory capturing the link of operators in the bulk and the higher-linking in the boundary.

\paragraph{SymTFT analysis.} We start by reminding the reader that the six dimensional $BF$-theory of level $n$ has an action given by
\begin{align}
    S=\frac{n}{2 \pi} \int_{M_6} b_k \wedge \diff c_{5-k} \, ,
\end{align}
where $b_k$ and $c_{5-k}$ are $k$-form and $(5-k)$-form $\U(1)$ connections respectively and $M_6$ is a six dimensional spin manifold. Provided the level $n$ is an integer, the theory is gauge invariant. The equations of motions are given by $n\diff b_k =0$ and $n\diff c_{5-k}=0$ which forces the two connections to become $\mathbb{Z}_n$ cocycles (i.e. $n b_k= \diff\lambda$). We can now define the gauge invariant Wilson operators
\begin{align}
    W_e(A)=e^{\I e \int_A b_k} \quad V_m(B)=e^{\I m \int_B c_{5-k}} \, ,
\end{align}
for $A$ and $B$ closed $k$-dimensional and $(5-k)$-dimensional manifolds respectively. These operators have a non trivial linking with one another. This can be easily seen by computing their expectation value
\begin{align} 
    \langle W_e(A)V_m(B)\rangle= e^{2 \pi \I \; (e \cdot m) \frac{ \mathrm{Link}_6(A,B)}{n}}.
\end{align}
Now, when the $BF$-theory admits a dynamical boundary, i.e. a $5$-dimensional field theory, we can describe the global structures of such theory in terms of the Wilson operators of the six dimensional $BF$ one.

To connect the $BF$-theory above to a field theory in one dimension lower, we now consider placing it on a $6$-manifold with two boundaries, homeomorphic to $[0,1] \times M$ where $[0,1]$ is the closed unit interval and $M$ is a $5$-dimensional manifold. Along the $M \times \{0\}$ boundary we impose a Dirichlet boundary condition (DBC) for $b_k$ or $c_{5-k}$ and on the other boundary we consider placing a QFT $\mathcal T_5$ with a $\mathbb Z_n^{(k)}$ symmetry. This picture can be summarised as follows:
\begin{align}
   \text{Dirichlet}: \quad\quad \langle D(B_k) |&=\sum_{b_k} \langle b_k| \delta(b_k-B_k), \\
   \text{Dynamical}: \hspace{0.6em}\quad\quad  | \cT_{5}\rangle &= \sum_{b_k } Z_{\cT_5}(M, b_k) |b_k \rangle.
\end{align}
where the sum is understood over all possible configuration of $b_k$ seen as a cocycle on the boundary geometry $M$ and $B_k$ is a fixed background for the $\mathbb Z_n^{(k)}$-form symmetry. As the $BF$-theory is topological, we can contract the interval $I$ so that the two boundaries coincide. This amounts to fixing a global structure for the field theory and providing an absolute theory in the sense of \cite{Freed:2012bs}. The partition of this absolute theory is then given by
\begin{gather}
    \langle D(B_k) | \cT_{5}\rangle = \sum_{b_k} \sum_{b'_k} \delta(b_k-B_k)Z_{\cT_5}(\cM_{D},b'_k)\langle b_k|b'_k\rangle = Z_{\cT_5}(\cM_{D},B_k)
\end{gather}
By choosing a DBC for $b_k$, this allows the $W_e(A)$ defects to end on the boundary, becoming a heavy defect in the dynamical theory. On the other hand, the $V_m(B)$ defects cannot end on the boundary, thus remaining a topological operator linking with the Wilson line. Due to the $BF$-theory's algebra, these operators therefore become the generators of the $k$-form symmetry acting on heavy defects. The lattice of all possible structure is spanned by the line operators of the $5$-dimensional theory. Choosing a maximal isotropic sub-lattice amounts in choosing a maximal commuting, i.e. with trivial linking, subset of lines.

In order to see how the above anomaly is captured by the expectation values of operators of the symmetry theory we will now consider the following 6d theory on a compact spin manifold $M_6$
\begin{align}\label{6dbbb}
    \mathcal{L} = \frac{n}{2 \pi} c_3 \wedge\diff b_2 + \frac{k}{6 \pi} b_2 \wedge b_2 \wedge b_2 \, .
\end{align}
The coefficient $k$ is fixed by requiring invariance under large gauge transformation and it is quantized according to flux identification \cite{Wang:2014pma}. This is achieved by considering a large gauge transformation for the $b_2 \to b_2 + \diff \beta_1$ field and requiring the action to be an integer multiple of $2 \pi$ with the condition $\int b_2 = 2 \pi \frac{l}{n}$ and $\int \diff \beta_1 = 2 \pi m$, for $l,m,n \in \mathbb{Z}$.

Under these transformations, the shift proportional to $k$ is given by
\begin{gather}
    \delta \cL \propto \frac{k}{6\pi}\left(3b_2\wedge b_2 \wedge\diff \beta_1 +3 b_2 \wedge \diff \beta_1 \wedge \diff \beta_1 + \diff \beta_1 \wedge\diff\beta_1 \wedge\diff\beta_1 \right).
\end{gather}
Integrating over spacetime and imposing $\delta \cL$ to be a multiple of $2\pi$, gives us\footnote{\ Note that here we use the fact that on spin manifolds $\int \diff \beta\wedge \diff \beta \wedge\diff \beta=6 (2 \pi)^3 m^3$, $\int \diff \beta \wedge\diff \beta=2 (2 \pi)^2 m^2$ and $\int b_2 \wedge b_2= 2(2 \pi)^2 l^2/n^2$ with $l,m,n \in \mathbb{Z}$.}
\begin{gather}
    \frac{k}{6 \pi} \left(\frac{6l^2 m}{n^2}(2 \pi)^3 +  \frac{6 l m^2}{n}(2 \pi)^3 + 6 m^3 (2 \pi)^3   \right)\, = 0 \quad (\mathrm{mod}\, 2\pi).
\end{gather}
This condition can be easily satisfied by taking 
\be
k=s n^2/4\pi
\ee
for $s \in \bbZ$.

Flux quantization imposes then a periodicity for $k$, this is simply given by imposing $b_2=\diff\lambda_1 / n$ in the action and we get
\begin{align}
\begin{gathered}
    \frac{s n^2}{6 (2\pi)^2} \int b_2 \wedge b_2 \wedge b_2=
    2\pi \frac{s l^3}{n} \, ,
\end{gathered}
\end{align}
which means that $k$ is an integer mod $n$.

Because of the anomaly in 5D, we know that we cannot gauge the 1-form symmetry for generic values of $k$. From the point of view of the symmetry theory, this is captured by the hypersurface topological operators and how they braid. To see this let us proceed in steps. First we need to check that the gauge invariance of the Lagrangian \cref{6dbbb}. It is easy to see that for $k \neq 0$ shifting $b_2$ and $c_3$ by exact forms does not maintain gauge invariance. Instead, in order to ensure gauge invariance we must allow for a more general gauge transformation for $c_3$
\begin{align}
    b_2 \to b_2 + \diff \beta_1 \, , \quad c_3 \to c_3 + \diff \gamma_2 -  \frac{2 k}{n} b_2 \wedge \beta_1 - \frac{k}{n} \beta_1 \wedge \diff \beta_1 \, , 
\end{align}
that leaves the Lagrangian invariant up to total derivatives. Despite the gauge invariance of the action, the line operator $V_m(A) = \exp\left( \I m \int_A c_3\right)$ is no longer gauge invariant under this more general transformation. This can be remedied by stacking an appropriately chosen TFT along $A$ in order to reabsorb the new gauge transformation
\begin{align}
    \tilde {V}_m(A) = \int {[\diff\phi]}\, e^{\I m \int_A (c_3 + \frac{s n }{4 \pi} \phi \diff \phi + \frac{s n}{2 \pi} \phi \wedge b_2)} \, ,
\end{align}
where $\phi$ is a 1-form defined on $A$, over which we are integrating\footnote{\ The operation of `stacking a TFT' to make operator gauge invariant is well know in the context of condensed matter physics, where the same operation is performed using inverse differential operators, see \cite{Zhang:2020kgc, Zhang:2021ycl,Putrov:2016qdo}.}. The new operator $\tilde {V}_m(A)$ is gauge invariant provided we transform $\phi \to \phi + \beta_1$.

The operation of stacking a TFT on an operator has the effect that the operator now obeys a non-invertible fusion rule. Explicitly, we have
\begin{align}\label{fusion-rule}
\begin{split}
    \tilde {V}_m(A) \times \tilde {V}_{-m}(A) &= \int {[\diff\phi][\diff\psi]} \,\exp\left(\I m \int_A \frac{s n}{4 \pi}\phi \diff \phi + \frac{s n}{2 \pi} \phi  b_2 - \frac{s n}{4 \pi }\psi \diff \psi - \frac{s n}{2 \pi} \psi  b_2 \right) 
    \\
    & =  \int {[\diff\phi][\diff\psi]}\,  \exp\left(\I m \int_A \frac{s n}{4 \pi}\alpha \diff \alpha + \frac{s n}{2 \pi} \alpha \diff \psi - \frac{s n}{2 \pi} \alpha  b_2 \right),
    \end{split}
\end{align}
where $\alpha=\phi-\psi$. Integrating out $\psi$ forces $\alpha$ to be a cocycle. Then, the first term vanishes, while the third term gives a sum of $W_e(\mathrm{PD}(\alpha))$ operators inserted on the Poincaré dual cycles of $\alpha$. This shows that the $\tilde {V}_m(A)$ operators are non-invertible operators in the symmetry theory.

We focus now on the correlator with the insertion of three operators $\tilde {V}_m(A)$. Explicitly
\begin{align}
    \langle\tilde {V}_{a_1}(A_1) \tilde {V}_{a_2}(A_2) \tilde {V}_{a_3}(A_3)\rangle = \int {\prod_i[\diff\phi]_i[\diff b_2][\diff c_3]} e^{\I\sum_i a_i \int_{A_i} (c_3 + \frac{s n }{4 \pi} \phi_i \diff \phi_i + \frac{s n}{2 \pi} \phi_i b_2) +\I S[b_2,c_3]}.
\end{align}
We now impose the equation of motion for $c_3$, enforcing 
\begin{align}\label{eq:c3motion}
    \diff b_2=2 \pi \sum_i \frac{a_i}{n} \mathrm{PD}(A_i),
\end{align} 
where $\mathrm{PD}(A_i)$ is the Poincaré dual of $A_i$. This can be solved considering bounding cycles $\partial \tilde{A}_i = A_i$, giving
\begin{align}
    b_2=2 \pi \sum_i \frac{a_i}{n} \mathrm{PD}(\tilde{A}_i).
\end{align} 

Imposing now the equation of motion for $\phi_i$ leads to 
\begin{align}\label{eq:phimotion}
    \diff \phi_i = b_2  \, ,
\end{align}
which must be satisfied on $A_i$. From \cref{eq:c3motion}, we know that $b_2$ cannot be exact, thus a solution to \cref{eq:phimotion} is possible only if the geometrical constraint
\begin{align}\label{eq:constraint}
    \mathrm{PD}(A_i) \wedge \left( \sum_j \frac{a_j}{n} \mathrm{PD}(\tilde A_j) \right) = 0 \, , 
\end{align}
is satisfied. This constraint forces the $A_i$ to not link each other in a pairwise fashion, so that the link is genuinely a 3-link.

Finally, we can substitute $b_2$ in the action $S[b_2, c_3]$, which leads to 
\begin{align}
    \mathrm{exp}\left( 2 \pi \I \frac{k}{6n} \left(\sum_i \frac{a_i}{n} \mathrm{PD}(\bar{A}_i)\right)^3 \right) \, ,
\end{align}
where we have now chosen $\partial \bar{A}_i = A_i$ such that \cref{eq:constraint} is satisfied. We now make the choice of a trivial framing\footnote{\ This theory can suffer from framing anomalies, the choice of trivial framing is consistent with the one in \cite{Kaidi:2023maf}. We leave a better understanding of the relation between framing of the operators in the symmetry theory and anomalies of the underlying field theory to a future work.} for the cycles $\tilde{A}_i$, i.e. their self intersection is trivial $\mathrm{PD}[\tilde{A}_i] \wedge \mathrm{PD}[\tilde{A}_i] = 0$. Upon expanding the above formula we get
\begin{align}\label{triple-intersec}
    \langle\tilde {V}_{a_1}(A_1) \tilde {V}_{a_2}(A_2) \tilde {V}_{a_3}(A_3)\rangle \approx e^{2 \pi \I \frac{s}{n} (a_1 \, a_2 \, a_3) \mathrm{Link}_{6}(A_1,A_2,A_3)} \, ,
\end{align}
where $\mathrm{Link}_{6}(A,B,C)$\footnote{\ A concrete realisation of this link as an embedding $\iota:S^3\sqcup S^3 \sqcup S^3\rightarrow\bbR^6$ is given in \cite{borromore}.} is an integer that encodes the triple linking of the three volumes $A_i$ in $M_6$, given by the triple intersection $\int \mathrm{PD}[\bar{A}_1] \wedge \mathrm{PD}[\bar{A}_2] \wedge \mathrm{PD}[\bar{A}_3]$. This corresponds to the triple linking number of type 0 defined in \cite{Kaidi:2023maf}, and it is an analogue in 6d of the Borromean configuration shown in \cref{fig:borro}.

As pointed out in \cite{Kaidi:2023maf}, this behaviour of the correlation function is related to an obstruction to gauging, i.e. to impose Neumann boundary conditions for the symmetry associated to the $b_2$ field. This result can also be interpret as an obstruction to the global variant realising the would be non-invertible 2-form magnetic symmetry. Let us also notice that the coefficient appearing in the exponent is trivial for any choice of charges $(a,b,c)$ when $s$ is a multiple of $n$. Moreover, if we chose $(a,b,c)$ so that the associate operators generate a subgroup of the 1-form symmetry, if $a \times b \times c$ is a multiple of $n$, the anomaly for that subgroup is trivial and that symmetry can be gauged.

Finally, we can notice that choosing
\be
s=\frac{q p(p-1)(p-2)}{n^2}\, \qquad n = \gcd(p,q)
\ee
we match our result from geometric engineering in equation \eqref{eq:triplegeomen}. Moreover, upon choosing DBC for $b_2$, correctly reproduces the 't Hooft anomaly of the 5D SCFT of \cite{Gukov:2020btk} as well.  Notice that the triple linking coefficient trivialises, precisely when the 't Hooft anomaly is trivial, i.e. when \cref{anomaly-condition} is satisfied. Thus, we can say that higher multi-linking are indeed obstructions to would be global structures. From this perspective, the defect groups correspond to the topological operators and defects in the SymTFT. In the above example, some of these operators are non-invertible and hence lack Dirichlet boundary conditions.

\section{Conclusions and outlook}\label{sec:conclusions}
In this work we have discussed how the geometric engineering dictionaries are enriched once one takes into account the global structures of the field theories involved. In particular, we have described an alternative construction of a \textit{geometric engineering at infinity} procedure, to characterize a $(D+1)$-dimensional bulk theory $\mathcal F_X$, responsible for these effects. We have tested our proposal in some key examples where we have reproduced previously established field theoretical results from a dual geometric engineering perspective.

\medskip

This work opens several directions to explore. First and foremost, in this note we do not discuss many details of the continuous symmetry cases that have been the subject of several recent studies \cite{Brennan:2024fgj,
Antinucci:2024zjp,
Bonetti:2024cjk,
Apruzzi:2024htg
}. We plan to return to this topic from the geometric engineering perspective advocated here in the near future. Secondly, in an upcoming work \cite{DMDZM}, we exploit this proposal to study more explicit examples with lower supersymmetry in the context of $G_2$ manifolds and Calabi-Yau 4-fold singularities, uncovering several new features of these backgrounds. In particular, recently, new $G_2$ geometries have been constructed \cite{Acharya:2023bth, Braun:2023fqa} that can be used as a fruitful playground to study more general features of the symmetry theories $\mathcal F_X$. Whenever $\mathcal F_X$ has sectors that are not purely topological, the symmetry inheritance mechanism of \cite{Acharya:2023bth} gives rise to a collection of bulk-boundary effects that are particularly interesting. Another direction we leave to investigate is the refinement of our construction as a function of framing of the link, and the resulting interplay with fractionalization, as introduced in \cite{Barkeshli:2014cna}. As a final remark, the interplay between stringy-dualities and non-invertible defects we discussed in this work gives a clear pathway to construct further examples of non-invertible defects in higher dimensional field theories, exploiting fiberwise selfdualities of more general backgrounds.

\section*{Acknowledgments}
We thank Bobby Acharya, Jonathan Heckman, Iñaki Garc\'ia Etxebarria, and Xiao-Gang Wen for discussions. MDZ thanks in particular Dan Freed, David Jordan, Ibou Bah, and Nytia Kitchloo, as well as Federico Bonetti and Ruben Minasian for discussions on closely related ongoing projects. In particular, a different more refined mathematical model for geometric engineering at infinity is currently being developed in a collaboration driven by David Jordan. MDZ also acknowledges a discussion with Fabio Apruzzi at the Simons Collaboration on Global Categorical Symmetries in November 2023. SM thanks Elias Riedel G\aa rding and Azeem Hasan for discussions and clarifications. The work of MDZ has received funding from
the European Research Council (ERC) under the European Union’s Horizon 2020
research and innovation program (grant agreement No. 851931). MDZ and SM also acknowledge support from the Simons Foundation (grant \#888984, Simons Collaboration on Global Categorical Symmetries). RM is supported by a Knut and Alice Wallenberg postdoctoral scholarship in mathematics and was supported in part by DOE grant DE-SC1019775 during the earlier stages of this project.

\appendix

\section{Higher linking numbers -- lightning review}\label{app:torsionallinks}

Link invariants have been deeply studied both in physics and mathematics -- see \cite{ massey1,highermassey,Horowitz:1989km,MR1990571,Witten:1988hf, PhysRevLett.113.080403,applications_mass,Wang:2014oya,Putrov:2016qdo,PhysRevB.96.085125,Chan:2017eov, Zhang:2021ycl} and reference therein. In this appendix we present a very quick review of the linking numbers used in the main text (with no pretense of rigor).

\paragraph{Non-torsional case.} Consider two closed oriented compact submanifolds of $\mathbb R^{d}$, denoted $\Sigma_1^{p_1}$ and $\Sigma_2^{p_2}$, of dimensions $\dim \Sigma_i^{p_i} = p_i$ such that $p_1 + p_2 + 1 = d$. Their linking number is defined as the degree of the map
\be
\varphi: \Sigma_1^{p_1} \times \Sigma_2^{p_2} \to S^{d-1} \qquad \varphi(x,y) = \frac{x-y}{|x-y|}.
\ee
An alternative to the above formula, which holds when the various ingredients below are well defined in a $d$-dimensional manifold $X$, is given by the following formula
\be\label{eq:link2}
\mathrm{Link}( \Sigma_1^{p_1}, \Sigma_2^{p_2}) = \int_X \mathrm{PD}(\widehat{\Sigma}_1^{p_1}) \wedge \diff \mathrm{PD}(\widehat{\Sigma}_2^{p_2}) = \widehat{\Sigma}^{p_1}_1 \cap \Sigma^{p_2}_2
\ee
where $\widehat{\Sigma}^{p_i}_i$ is a $p_i+1$ dimensional Seifert surface such that $\partial \widehat{\Sigma}^{p_i}_i = \Sigma^{p_i}_i$ and $\mathrm{PD}$ is the Poincaré duality map
\be
\int_\Sigma \alpha = \int_X \alpha \wedge \mathrm{PD}(\Sigma)
\ee
that to every compact $k$ dimensional cycle assigns a $(d-k)$-form. Of course, in manifolds with a non-trivial homology Seifert surfaces can be obstructed. This complicates the schematic definition of higher linking we are giving here.\footnote{\ A more precise treatment can be based on the theory of higher Massey products and $A_\infty$ structures.} Here we  assume that the various cycles we are considering are boundaries. Since we are interested only in certain specific higher links, we can choose to restrict the support of the various strands of our links to satisfy this extra assumption (by working in a small enough neighborhood). Notice that since the Seifert surfaces have one dimension higher, a necessary condition for equation \eqref{eq:link2} is precisely that $p_1 + p_2 + 1 = d$. For a link with $L$ strands in $d$ dimensions there are $L-1$ possible linking numbers, namely
\be\label{eq:lllink}
\mathrm{Link}^{(\kappa)}_d( \Sigma_1^{p_1}, \Sigma_2^{p_2},...,\Sigma_L^{p_L}) = \int   \mathrm{PD}(\widehat{\Sigma}_1^{p_1}) \wedge \dots \wedge  \mathrm{PD}(\widehat{\Sigma}_{L-\kappa}^{p_{L-\kappa}}) \wedge \diff \mathrm{PD}(\widehat{\Sigma}_{L-\kappa+1}^{p_{L-\kappa+1}})  \dots \wedge \diff  \mathrm{PD}(\widehat{\Sigma}_L^{p_L})
\ee
where $0 \leq \kappa \leq L-1$. Moreover, since we are interested in those configurations that highlight certain specific higher order correlators in the SymTFT, we will require that the $L$-links are Brunnian, meaning that any sub-collections of $1< \ell < L$ strands $\Sigma^{p_{i_1}}_{i_1},...,\Sigma^{p_{i_\ell}}_{i_\ell}$ out of the ones involved in \eqref{eq:lllink} are trivially linked, namely:
\be\label{eq:brunnian}
\mathrm{Link}_d^{(\kappa')}(\Sigma^{p_{i_1}}_{i_1},...,\Sigma^{p_{i_\ell}}_{i_\ell}) = 0
\ee 
for all $0< \kappa' < \ell-1$. In a torsionless compact $d$-dimensional space the higher linking numbers so defined are integers and the higher linking pairing has symmetry properties depending on $d$, $p_1$, ..., $p_L$ which can be inferred by integration by parts in the equation above. A necessary condition for non-trivial higher links is that the dimensions of the various supports involved satisfy:
\be\label{eq:higherlinkconstraint}
(L-1)d = L-\kappa + \sum_{i=1}^L p_i.
\ee
We stress that only if the links involved are Brunnian, the expressions in \eqref{eq:lllink} coincide with the homotopic higher link invariants. For non-Brunnian links, the expression we use for higher linking numbers receive corrections to cancel their dependence on Seifert surfaces. Often a good technique to detect such corrections is to exploit generalised Wilson lines in suitably defined topological field theories, see eg. \cite{Putrov:2016qdo} for some applications in $d=3,4$. This allows one to find the required modifications on a case by case analysis (but a general universal expression is not known to us). In this work, however, we are only interested in certain specific correlators in the SymTFT that can detect obstructions to certain global structures, and not to the most general expressions. For this purpose we can require the links involved to be Brunnian and neglect the dependence on higher linking on framing (as long as we neglect finer data such as the dependence on a choice of spin structure). It is interesting to remark that for non-Brunnian higher links with $L$ strands such that $\kappa = L-1$ the higher linking numbers we list above are automatically homotopic invariants. Remarkably these are the only ones that seem to contribute in the examples we have considered in this paper.

\paragraph{The case of torsional cycles.} For a $d$-dimensional manifold $X$, one can also have torsional $p_i$-cycles, i.e. cycles of dimension $p_i$ such that there is a $(p_i+1)$-dimensional cycle $\gamma_i$ with the property 
\be
\partial \gamma_i^{p_i+1} = K_i \beta^{p_i}
\ee
where $K_i\in \mathbb Z_{>1}$. The cycles $\gamma_i$ then play the roles of the Seifert surfaces in our definitions above and can be used to extend the notion of higher linking numbers among torsional cycles, e.g. for $0 \leq \varkappa \leq L-1$ we have
\be\label{eq:torsionhigherlink}
\begin{aligned}
&\mathrm{Link}_d^{(\varkappa)}( \beta_1^{p_1}, \beta_2^{p_2},...,\beta_L^{p_L}) =\\
&\, = {1 \over K_1 K_2 ... K_{L}} \int_X   \mathrm{PD}(\gamma_1^{p_1+1}) \wedge \dots \wedge  \mathrm{PD}(\gamma_{L-\varkappa}^{p_{L-\varkappa}+1}) \wedge \diff \mathrm{PD}(\gamma_{L-\varkappa+1}^{p_{L-\varkappa+1}+1})  \dots \wedge \diff  \mathrm{PD}(\gamma_L^{p_{L}+1})
\end{aligned}
\ee
where we also require equation \eqref{eq:brunnian} to hold. In this case, the resulting higher linking numbers are rational. Of course, the above definition includes the former one: non-torsional cycles are such that $K_i = 1$, in particular the same restriction on dimensions (equation \eqref{eq:higherlinkconstraint}) applies.


\subsection{The case of \texorpdfstring{$\mathbf{L}_X = T^2 \times S^3/\mathbb Z_N$}{LX = T2 x S3/ZN}}\label{app:KK4DN=4} 

\begin{table}
\begin{center}
\begin{tabular}{c||ccc||c}
$L$ & $x_1$ & $x_2$ & $x_3$ & \phantom{\Big|} $\varkappa$\phantom{\Big|} \\
\hline
2 & - & 2 & - &1\\
   &  1 & - & 1 &1 \\
     &1 & 1 & - & 0\\
\hline
3 & - & - & 3 & 2\\ 
 &  - & 1 & 2 & 1\\
 & - & 2 & 1 & 0 \\
 & 1 & - & 2 & 0\\
 \hline
 4 & -  &- & 4 & 1\\ 
  &- & 1 & 3 & 0 \\ 
 \hline
  5 & - & - & 5 & 0
\end{tabular}
\caption{Possible higher linkings on $L_X$}\label{tab:ella}
   \end{center}

\end{table}

In this appendix we discuss the details of the possible higher linkings of $\mathbf{L}_X = T^2 \times S^3/\mathbb Z_N$. This space has dimension 5, and torsional cycles of dimensions $1,2$ and $3$. 
Consider a link with $L$ strands consisting of $x_i$ strands of dimension $i$. The necessary condition on dimensions of cycles, Equation \eqref{eq:higherlinkconstraint}, gives in this example
\be\label{eq:nec1}
x_1 + 2 x_2  + 3 x_3 + L - \varkappa = (L - 1) 5 \qquad\text{and}\qquad x_1 + x_2 + x_3 = L
\ee
where $0 \leq \varkappa \leq L-1$. These equations have a finite number of solutions that we list in Table \ref{tab:ella}. These are the possible higher links one can construct between the torsional cycles of $\mathbf{L}_X$. Corresponding to each entry of the table one should check there is a compatible link for the bulk 5-dimensional symmetry theory from geometric engineering at infinity. Given a solution $(x_1,x_2,x_3;\varkappa)$ we consider the IIA $\mathbf{Dp}_a$ branes that can give rise to links of $L$ strands consisting of $y_{a,i}$ strands that are $(p_a + 1 - i)$-dimensional. The necessary condition for such links to exist is that for each triple solving \eqref{eq:nec1} also the following equations have solutions:
\be\label{eq:sbam}
\sum_{i=1}^3 \sum_{a} y_{a,i} (p_a + 1 - i)  + L - \kappa = (L - 1) 5 \qquad  x_i = \sum_a y_{a,i} 
\ee
for $0 \leq \kappa \leq L-1$ and $0< (p_a + 1 - i) < d$. In our case $d=4$. The relevant branes in IIA are D2 branes on torsional 1-cycles and 2-cycles and D4 branes on torsional 2-cycles and 3-cycles.

Consider for example the second entry in the table, this gives rise to a the electric and magnetic 1-form symmetries of the model as we discussed in the main body of the text. To find possible obstructions to gauging one or the other, we need to look at the table for possible higher linking involving the torsional 3-cycles or the torsional 1-cycles. Consider for example the torsional 3-cycles. The possible higher linking that involve these and the 1-cycles are provided by $(x_1,x_2,x_3;\varkappa) = (0,0,3;2), (0,0,4;1), (0,0,5;0)$ and $(1,0,2;0)$. It is easy to see that in each case there is no solution with a positive $\kappa$, 
and hence there are no obstructions to gauge the magnetic one-form symmetry. Consider as an example the case $(0,0,3;2)$. Here we can wrap D4 branes on the three 3-cycles and form a type 2 link with $L=3$ strands along $\mathbf{L}_X$. This configuration gives $y  = 3$, and the coresponding cycles have dimension $p+1-3 = 2$, $L=3$ and the resulting equation for $\kappa$ is $3 \times 2 + 3 - \kappa = 2 \times 5$ which implies $\kappa = -1$. Hence we discard this configuration. The other 3 cases are analyzed similarly. The case $(1,0,2;0)$ gives an example with two species of branes involved, namely D4s along the 3-cycles and D2s along the 1-cycles. In this case we have $y_{D2,1} = 1$ and $y_{D4,3} = 2$ and the corresponding topological defects with 2-dimensional supports for which the same constraint applies.

We stress here that there are more higher symmetries for the theory $T^2 \times \mathbb C^2/\mathbb Z_N$ than one would naively expect from looking at 4D $\mathcal N=4$ SYM. This is a consequence of the fact that this geometry gives a KK theory for the 6d $\mathfrak{su}_N$ (1,1) theory on $T^2$, indeed the latter is obtained from geometric engineering IIA on $\mathbb C^2/\mathbb Z_N$ and our geometry is simply $T^2 \times \mathbb C^2/\mathbb Z_N$. For example, in the 5D SymTFT from $T^2 \times S^3/\mathbb Z_N$ we expect to find line operators corresponding to D2 branes wrapping torsional 2-cycles as well as 3-surface operators corresponding to D4 branes wrapping on them. Corresponding to the linking $(0,2,0;1)$ in Table \ref{tab:ella} we have that for $y_{D4,2} = 1$ and $y_{D2,2} = 1$ the equation \eqref{eq:sbam} has an interesting solution, indeed $1 \times 3 + 1 \times 1 + 2 - \kappa = 5$ gives $\kappa = 1$. This gives rise to a Heisenberg algebra of topological membranes
\be
\langle \mathcal D_{\mathbf{D4}}^\beta(\Sigma^3)\mathcal D_{\mathbf{D2}}^{\beta'}(\Sigma^1)\cdots\rangle = \text{exp}\left(2\pi \I \, {\beta \beta' \over N} \mathrm{Link}_5(\Sigma^3,\Sigma^1)\right) \langle \cdots \rangle
\ee
that captures the remant of the defect group of the 6d $\mathfrak{su}_N$ (1,1) theory, namely $(\mathbb Z_N)^{(1)}_e \oplus (\mathbb Z_N)^{(3)}_m$ reduced on $T^2$. We expect that only the KK modes are charged with respect to these extra higher symmetries, in order to focus on the 4D $\mathcal N=4$ theory we can discard the topological defect operators that arise from reductions of D2 and D4 on torsional 2-cycles.

\subsection{The case of \texorpdfstring{$X = \text{Cone}(Y^{p,q})$}{X=Cone(Yp,q)}}\label{app:Ypq}

In this appendix we discuss in detail the geometry relevant for the analysis of the $\SU(N)_k$ example in Section \ref{sec:ex5d}. Consider the case of torsional cycles in $\beta^1 \in \text{Tor } H_1(Y^{p,q})$ and $\beta^3 \in \text{Tor } H_3(Y^{p,q})$. For both cases we expect that $\gcd(p,q) \beta^{k} = \partial \gamma^{k+1}$ for $k=1,3$. We can use the same analysis as in the previous section, indeed the solutions to
\be
x_1 + 3 x_3 + L-\varkappa = (L-1)5 \qquad x_1 + x_3 = L
\ee
correspond to the $x_2 = 0$ solutions in Table \ref{tab:ella}. In this context Equation \eqref{eq:sbam} generalises to
\be\label{eq:sbam2}
\sum_{i=1}^3 \sum_{a} y_{a,i} (p_a + 1 - i)  + L - \kappa = (L - 1) 6 \qquad  x_i = \sum_a y_{a,i} 
\ee
for $0 \leq \kappa \leq L-1$ and $0< (p_a + 1 - i) < d$. In this case $d=5$. The only cases of interest are M5s on torsional 3-cycles and M2s on torsional 1-cycles (other topological configurations are either too big or too small). These correspond to the topological operators
\be
\mathcal D^{\beta^1}_{\mathbf{M2}}(\Sigma^2) \qquad \mathcal D^{\beta^3}_{\mathbf{M5}}(\Sigma^3)
\ee 
The solutions $(x_1,x_3;\varkappa) = (1,1;1)$ corresponds to the Heisenberg algebra in Equation \ref{eq:Heisenbergo5d}. The higher linkings $(x_1,x_3;\varkappa) =(1,2;0),(0,3;2),(0,4;1),(0,5;0)$ need to be discussed. Of these the only solution with a positive $\kappa$ is the case $(0,3;2)$ for which \eqref{eq:sbam2} reads:
\be
3 \times 3 + 3 -\kappa = 2 \times 6 \quad \Rightarrow \quad \kappa = 0
\ee
and hence we obtain a Borromean type link among three $S^3$ in the bulk 6-dimensional topological theory --- see Figure \ref{fig:borro}. All the other give rise to solution with a negative $\kappa$ that must be discarded, for example if we consider the case $(0,5;0)$ we obtain the equation $5 \times 3 + 5 - \kappa = 4 \times 6$ that has no solutions for $\kappa \geq 0$.

\newpage

\phantomsection
\addcontentsline{toc}{section}{References}
\bibliography{6d_bib_plus_plus}{} 
\end{document}